\documentclass[preprint,secnumarabic,tightenlines,amssymb, amsmath,nobibnotes,
aps,nofootinbib,showpacs,]{revtex4-1}%
\usepackage{graphicx}
\usepackage[caption=false]{subfig}
\usepackage{dcolumn}
\usepackage{bm}
\usepackage{amssymb}
\usepackage{setspace}
\usepackage{amsmath}
\usepackage{amsfonts}
\usepackage{verbatim}
\usepackage{MnSymbol}
\usepackage{hyperref}
\usepackage{color}
\usepackage{inputenc}
\DeclareMathOperator{\sgn}{sgn}
\providecommand{\U}[1]{\protect\rule{.1in}{.1in}}
\definecolor{darkgreen}{rgb}{0,0.35,0}

\definecolor{Rood}{rgb}{1, 0, 0}
\begin{document}
\title{\noindent\textbf{Comments on the compatibility of thermodynamic equilibrium conditions
with lattice propagators}}
\author{Fabrizio Canfora${}^{1}$}
\thanks{fcanforat@gmail.com}
\author{Alex Giacomini${}^{2}$}
\thanks{alexgiacomini@uach.cl}
\author{Pablo Pais${}^{1,3}$}
\thanks{pais@cecs.cl}
\author{Luigi Rosa${}^{4,5}$}
\thanks{rosa@na.infn.it}
\author{Alfonso Zerwekh${}^{6}$}
\thanks{alfonso.zerwekh@usm.cl}
\affiliation{$^{1}$ Centro de Estudios Cient\'{\i}ficos (CECs), Casilla 1469, Valdivia, Chile}
\affiliation{$^{2}$ Instituto de Ciencias F\'{i}sicas y Matem\'{a}ticas, Universidad Austral de Chile, Casilla 567, Valdivia, Chile}
\affiliation{$^{3}$ Physique Th\'{e}orique et Math\'{e}matique, Universit\'{e} Libre de
Bruxelles and International Solvay Institutes, Campus Plaine C.P.~231, B-1050
Bruxelles, Belgium}
\affiliation{$^{4}$ Dipartimento di Fisica, Universit\'{a} di Napoli Federico II, Complesso
Universitario di Monte S.~Angelo, Via Cintia Edificio 6, 80126 Napoli, Italia}
\affiliation{$^{5}$ INFN, Sezione di Napoli, Complesso Universitario di Monte S.~Angelo,
Via Cintia Edificio 6, 80126 Napoli, Italia}
\affiliation{$^{6}$ Departamento de F\'{\i}sica and Centro Cient\'{\i}fico-Tecnol\'{o}gico
de Valpara\'{\i}so Universidad T\'{e}cnica Federico Santa Mar\'{\i}a Casilla
110-V, Valpara\'{\i}so, Chile}

\begin{abstract}
In this paper the compatibility is analyzed of the non-perturbative equations of state of quarks and gluons arising from the lattice with some natural requirements for self-gravitating objects at equilibrium: the existence of an equation of state (namely, the possibility to define the pressure as a function of the energy density), the absence of superluminal propagation and Le Chatelier's principle.
It is discussed under which conditions it is possible to extract an equation of state (in the above sense) from the non-perturbative propagators arising from the fits of the latest lattice data. In the quark case, there is a small but non-vanishing range of temperatures in which it is not possible to define a single-valued functional relation between density and pressure. Interestingly enough, a small change of the parameters appearing in the fit of the lattice quark propagator (of around 10~\%) could guarantee the fulfillment of all the three conditions (keeping alive, at the same time, the violation of positivity of the spectral representation, which is the expected signal of confinement). As far as gluons are concerned, the analysis shows very similar results. Whether or not the non-perturbative quark and gluon propagators satisfy these conditions can have a strong impact on the estimate of the maximal mass of quark stars.
\end{abstract}
\maketitle

\section{Introduction}

One of the main open problems in theoretical physics is a proper understanding
of the infrared behavior of non-Abelian gauge theories, like Quantum
Chromodynamics (QCD), and of its phase diagram (see \cite{greenconf}). The
non-perturbative nature of the infrared region of QCD prevents one from using
the standard perturbative techniques based on Feynman diagrams. Thus, it is
necessary to rest on non-perturbative techniques and/or lattice data. In the
present paper, we will combine lattice data (both for the quarks and gluons
propagators) together with the non-perturbative effects arising from (the
elimination of) Gribov copies \cite{Gribov:1977wm}\ (for the gluonic sector)
to extract non-perturbative equations of state for quarks as well as gluons.
We will adopt the $\zeta$-function regularization technique
\cite{brown:69,dowker:76,hawking:77}, which allows one to write many of the
physical quantities in a closed form as (very rapidly convergent) series of
Bessel functions. This technical point will play an important role in the
following.\newline

The details of non-perturbative propagators of quarks and gluons are of great
interest in applications. For instance, in astrophysics, there is evidence
supporting the existence of quark stars (two detailed reviews are
\cite{quarkstar1,quarkstar2}). As such objects are gravitating, it
is of fundamental importance to check under which conditions they can be in
hydrostatic equilibrium and which is the expected upper bound on their mass.
Due to the great difficulty to construct analytically an equation of state
(EOS) for such extreme matter configurations, it is important to have some
estimates on the mass bounds of self-gravitating objects which can be deduced
by generic principles rather than from the exact form of the EOS. For a
neutron star this has been done by Rhoades and Ruffini \cite{ruffini} using
only very basic principles. Two of the required basic principles are the
absence of superluminal propagation and Le Chatelier's principle. \newline

The absence of superluminal propagation is necessary to enforce causality and
is considered one of the most basic principles of relativistic physics.

Le Chatelier's principle simply states that when any system at equilibrium is
subjected to change, it will react in such a way to oppose to the change.
This means that assuming Le Chatelier's principle is completely generic as,
without it, it would not even be possible have a stable equilibrium.

The third principle required by Rhoades and Ruffini is the validity of the
hydrostatic equilibrium equation of General Relativity. Just from the above
requirements, they were able to derive a bound of $3.2$ $M_{\bigodot}$ for a
neutron star without knowing any further details of the EOS (besides its
existence of course, as it will be explained in a moment).\newline

As it is possible to extract detailed information as regards on the non-perturbative
propagators from lattice data (as well as from the Gribov-Zwanziger procedure)
a natural question arises: \textit{will the equations of state derived from
the non-perturbative quarks and gluons propagators be compatible with the
basic principles stated above}?

In the present analysis, in order to be as generic as possible, we will drop
the assumption of the hydrostatic equilibrium equation as it is specific to
General Relativity (it may be that this theory acquires some corrections in
some extreme range of parameters). The other two principles, namely causality
and Le Chatelier's principle are, instead, quite generic and, consequently, we
will consider them.

Actually, before discussing these two principles, there is a more basic
requirement which was implicitly assumed in \cite{ruffini} (as well as in a
great part of the theoretical literature on gravitating compact objects):
namely the \emph{existence} of a well-defined EOS or, in other words, the
possibility to define an implicit one-to-one functional relation between pressure
and energy density. The importance of such a requirement becomes obvious if
one considers that, if it is not satisfied, it is not even possible to discuss
the coupling with the Einstein equations (in the usual way, at least). That is
the reason why in \cite{ruffini} such a principle was not analyzed in detail,
rather it was just considered as obvious. However, the present analysis shows
that the very existence of an EOS depends on the precise values of the
parameters appearing in the fit of the lattice propagators. Remarkably, there
are cases in which it is not possible at all to extract a well-defined EOS
from the non-perturbative propagators. This, for instance, prevents one from
coupling strongly interacting quarks and gluons to Einstein gravity in any
obvious way. In particular, in these situations the bound derived in
\cite{ruffini} would not apply. Usually, the fact that it is not possible to
define a one-to-one relation between pressure and energy density suggests that
some extra physical parameter is needed to properly label the equilibrium
states in order to define an EOS in the usual sense.\newline

The fact that there may not exist a well-defined EOS in certain situations can
be seen as follows. From the non-perturbative propagator it is possible to
compute the grand partition function from which all relevant thermodynamical
quantities like the pressure and energy density as functions of the
temperature and of the chemical potential can be derived. Taking, for example,
a fixed value of the chemical potential it is possible to get a parametric
equation for the energy density $e(T)$ versus pressure $P(T)$ curve. This
curve, however, represents a single-valued function $P=P(e)$ only when the
functions $P(T)$ and $e(T)$ are strictly monotonous as function of the
temperature $T$ or the shapes and ranges of non monotonicity in $T$ for $P$
and $e$ are exactly the same.

The computation of the grand partition function from the non-perturbative
lattice quarks propagator at finite temperature and chemical potential has
been performed in the reference \cite{quarkNP6} using dimensional
regularization techniques. In this paper, the $\zeta$-function regularization
technique is used instead and the non-perturbative gluons propagator is
discussed as well. The advantage of the $\zeta$-function regularization
\cite{actor:85,actor:86} is that it reduces the number of numerical integrations
(many of the expressions are evaluated as fast converging series of Bessel
functions) and, consequently, it reduces the numerical error allowing to see
even very tiny effects. Indeed, although our computations show that the two
techniques clearly agree, using the $\zeta$-function approach it is possible
to see that for the values of the parameters appearing in the fit of the last
lattice quarks propagator, the curve $P(T)$ and $e(T)$ of the pressure and
energy density as a function of the temperature are non-strictly monotonic
functions and have different shapes near $T=0$. This means that for a small range of temperatures, the functional relation of
pressure and energy density is not one-to-one: namely, one of the requirements
of Rhoades and Ruffini is violated.\newline

The problem in the EOS being well defined disappears by allowing a
change in the fit parameters of around $10~\%$. Interestingly, once the
existence of an EOS has been ensured, the causality and Le Chatelier
principles turn out to be almost satisfied without any extra
requirements on the parameters. For this reason, it may be interesting to
explore the phenomenological consequences of allowing such changes in the fit parameters.

A similar analysis reveals that the same requirement of Rhoades and Ruffini
can be violated in the case of the non-perturbative gluons EOS. The
theoretical interest of the gluon case is that if one considers a
Gribov-Zwanziger (GZ) propagator with the inclusion of the effects of the
condensates then, by allowing a change in the fit parameters of around $10~\%$,
one can satisfy all the requirements of Rhoades and Ruffini. On the other
hand, if one does not include the condensates, then a change in the Gribov
mass of around $10~\%$ is definitely not enough to satisfy all the requirements
of Rhoades and Ruffini.

The paper is organized as follows: in the next section the relevant
thermodynamic quantities for the non-perturbative quark propagator are
constructed. The most relevant details of the computations related to $\zeta
$-function regularization technique are kept in the main text as they are
fundamental in detecting the non-existence regime of the EOS. In the third
section, the analysis is extended to the gluonic sector. The last section is
dedicated to the conclusions and perspectives.

\section{The non-perturbative quark propagator and its thermodynamics}

Let us start the analysis with the non-perturbative quark propagator $S(p)$
arising from the lattice \cite{quarkNP1,quarkNP2}
\begin{align}
S\left(  p\right)   &  =-\frac{\gamma_{\mu}p_{\mu}+\mathbf{1}_{4}%
\mathbf{\ }M_{0}\left(  p\right)  }{p^{2}+M_{0}^{2}\left(  p\right)
}\ ,\ \ M_{0}\left(  p\right)  =\frac{M_{3}}{p^{2}+m^{2}}+m_{0}%
\ ,\label{quarkp1}\\
M_{3}  &  =0.196\ GeV^{3},m^{2}=0.639\ GeV^{2},m_{0}=0.014\ GeV\ ,
\label{quarkp2}%
\end{align}
where $\gamma_{\mu}$ are the Euclidean Dirac matrices and $\mathbf{1}_{4}$ is
the $4\times4$ identity matrix. Such a propagator can be included in the framework of
the so-called refined Gribov-Zwanziger (RGZ) approach as indicated in
\cite{quarkNP5,quarkNP5.5}. It is worth emphasizing here that the
above propagator can be expanded into three `standard' Fermions propagators,
two of which having complex conjugated poles and the third with real poles.
The physical interpretation of having complex conjugated poles is that they
are a signal of confinement as propagators with complex poles in $p^{2}$
violate positivity. Such poles are determined by the following equation:
\begin{equation}
p^{2}\left(  p^{2}+m^{2}\right)  ^{2}+\left[  M_{3}+m_{0}\left(  p^{2}%
+m^{2}\right)  \right]  ^{2}=\\
\left(  p^{2}+\alpha_{1}\right)  \left(  p^{2}+\alpha_{2}\right)  \left(
p^{2}+\alpha_{3}\right)  , \label{poles1}%
\end{equation}
meaning that the set $\{\alpha_{i}\}$ represents minus the roots of the
denominator in $S(p)$.

\subsection{Thermodynamics}

As said above, the idea to compute the grand partition function using the
non-perturbative quarks propagator arising from lattice data has already been
proposed in the reference \cite{quarkNP6}. The new idea proposed in the
present paper is to analyze in which range of the parameters appearing in the
propagators, pressure and energy density satisfy some very basic consistency
conditions of thermodynamical equilibrium configurations. Such conditions
appear in a very natural way in the analysis of any self-gravitating compact
object (like quark stars) in an analogous way to the pioneering work
\cite{ruffini}.

The first consistency condition is actually so obvious that it was not even
explicitly enumerated (but of course assumed) in \cite{ruffini} so that we
will call it \textit{condition zero}.

\textit{Condition zero: Existence of an equation of state} In the standard
general relativistic approach to self-gravitating objects, before even
beginning to require some consistency conditions on thermodynamical
quantities, it is necessary to have an EOS (namely, a functional relation
$P=P(e)$ of the pressure in terms of the energy density). If this condition is
not satisfied, there would not exist any obvious way to couple non-perturbative
quarks and gluons to General Relativity (nor to any reasonable generalization
of General Relativity). As there are some arguments supporting the existence
of quark stars \cite{quarkstar1,quarkstar2}, the issue about the possibility
to define an EOS even in such extreme conditions is very relevant.\newline

\textit{Condition one: Causality} As is well known, in order to enforce
causality a necessary condition is to impose the requirement that no signal can travel faster
than the speed of light. This means that the speed of sound inside a self-gravitating object cannot be superluminal. In terms of the EOS this condition
takes the simple form $\frac{\mathrm{d}P}{\mathrm{d}e}\leq1$. It is therefore obvious that,
without condition zero, it is not even possible to enforce causality.\newline

\textit{Condition two: Le Chatelier's Principle} This principle states that for any action intended to modify a
given equilibrium configuration of a system, the system will react in such a
way that it opposes to the change. This principle is actually equivalent to
the assumption that there exists a stable equilibrium configuration. For a
self-gravitating object this principle implies that there is no spontaneous
gravitational collapse. In terms of the EOS, this principle can be stated as
$\frac{\mathrm{d}P}{\mathrm{d}e}\geq0$. Once again, it is worth noticing here that without
condition zero being satisfied, it is not possible to enforce this
principle.\newline

The condition that is prone to fail in the context of the non-perturbative
quarks propagator (as well as gluons propagator in the next section) is
\textit{condition zero}. Indeed we will show that it exists a narrow range of
temperature close to $T=0$ where the pressure is not a strictly monotonous
function. Very important, from this point of view, is the fact that the
$\zeta$-function regularization provides results in closed analytic form as
sums of (very fast convergent) series of suitable Bessel functions (unlike the
reference \cite{quarkNP6} in which the authors used dimensional
regularization). The grand partition function corresponding to the
non-perturbative quark propagator in equations (\ref{quarkp1}) and (\ref{quarkp2})
reads (we will follow the notation of \cite{quarkNP6}):
\begin{equation}
\frac{\log Z(T,\mu)}{2\beta VN_{c}N_{f}} =\sum\limits_{n=-\infty}^{+\infty
}\int\frac{d^{3}p}{(2\pi)^{3}}\log\Lambda^{-2}\left( \Omega^{2}(\mu
)+(\omega_{n}-i\mu)^{2}\right)  , \label{partition_func1}%
\end{equation}
where in our case $\Lambda$ is a suitable dimensional parameter, $N_{c}=3$,
$N_{f}=6$, $\omega_{n}=2\pi(n+1)T$ are the Matsubara frequencies for fermions,
and
\begin{equation}
\Omega^{2}(\mu)=\bm{p}^{2}+\left(  \frac{M_{3}}{(\omega_{n}-i\mu
)^{2}+\bm{p}^{2}+m^{2}}+m_{0}\right)  ^{2}.
\end{equation}
It is worth to note that usually the grand partition function is written with
$\beta^{2}=1/T^{2}$ instead of our $\Lambda^{-2}$. However, some subtraction
must be done also in such a case in order to avoid the infinite part
\cite{Kapusta}. In our case, splitting our regulator as $\Lambda^{2}=\beta
^{2}\mathcal{N}^{2}$ in \eqref{partition_func1}, where $\mathcal{N}$ is a
dimensionless constant, we obtain
\[
\frac{\log Z(T,\mu)}{2\beta VN_{c}N_{f}} =\sum\limits_{n=-\infty}^{+\infty
}\int\frac{d^{3}p}{(2\pi)^{3}}\log\beta^{-2}\left( \Omega^{2}(\mu)+(\omega
_{n}-i\mu)^{2}\right)  + \sum\limits_{n=-\infty}^{+\infty}\int\frac{d^{3}%
p}{(2\pi)^{3}}\log\mathcal{N}^{-2}\; ,
\]
where we can use precisely the last term to absorb the infinity setting
$P(T=0,\mu=0)=0$. At the end, this will give us the same result as
\cite{quarkNP6}.

One can simplify the above expression \eqref{partition_func1} as
{\footnotesize
\begin{align*}
\frac{\log Z(T,\mu)}{2\beta VN_{c}N_{f}}  &  =\sum\limits_{n=-\infty}%
^{+\infty}\int\frac{d^{3}p}{(2\pi)^{3}}\ln\Lambda^{-2}\left[  \bm{p}^{2}%
+\left(  \frac{M_{3}}{(\omega_{n}-i\mu)^{2}+\bm{p}^{2}+m^{2}}+m_{0}\right)
^{2}+(\omega_{n}-i\mu)^{2}\right] \\
&  =\sum\limits_{n=-\infty}^{+\infty}\int\frac{d^{3}p}{(2\pi)^{3}}\ln
\Lambda^{-2}\left[  \frac{\left(  \bm{p}^{2}+(\omega_{n}-i\mu)^{2}\right)
\left(  \bm{p}^{2}+(\omega_{n}-i\mu)^{2}+m^{2}\right)  ^{2}+\left(
M_{3}+m_{0}\left(  \bm{p}^{2}+(\omega_{n}-i\mu)^{2} +m^{2} \right)  \right)
^{2} }{\left(  \bm{p}^{2}+(\omega_{n}-i\mu)^{2}+m^{2}\right)  ^{2}}\right] \\
&  =\sum\limits_{i=1}^{i=4}c_{i}\sum\limits_{n=-\infty}^{+\infty}\int
\frac{d^{3}p}{(2\pi)^{3}}\ln\Lambda^{-2}\left[  \bm{p}^{2}+\alpha_{i}%
^{2}(\omega_{n},\mu)\right]  =:\sum\limits_{i=1}^{i=4}I^{(\alpha_{i})}c_{i},
\end{align*}
} where $\left\{  \alpha_{i}^{2}(\omega_{n},\mu),~(i=1,2,3)\right\}  $ are
minus the three roots of the numerator, $\alpha_{4}^{2}=(\omega_{n}-i\mu
)^{2}+m^{2}$, $\left\{  c_{i}=1,~(i=1,2,3)\right\} $ and $c_{4}=-2$. So,
everything reduces to find for a generic $\alpha_{i}$, the quantity
\[
I^{(\alpha_{i})}=\sum\limits_{n=-\infty}^{+\infty}\int\frac{d^{3}p}{(2\pi
)^{3}}\ln\Lambda^{-2}\left[  \bm{p}^{2}+\alpha_{i}^{2}(\omega_{n},\mu)\right]
.
\]
%@@@@@@@@@@@

Now, using the fact that $\ln\Lambda^{-2}\left[  \bm{p}^{2}+\alpha_{i}%
^{2}(\omega_{n},\mu)\right]  =-\lim_{s\rightarrow0}\frac{\partial}{\partial
s}\ln\left(  \Lambda^{2}\left[  \bm{p}^{2}+\alpha_{i}^{2}(\omega_{n}%
,\mu)\right]  \right)  ^{-s}$ we can write:
\begin{align}
I^{(\alpha_{i})}  &  =-\lim_{s\rightarrow0}\frac{\partial}{\partial s}\left\{
\left(  {\Lambda}\right)  ^{2s}\sum\limits_{n=-\infty}^{+\infty}\int
\frac{d^{3}p}{(2\pi)^{3}}\left(  \left[  \bm{p}^{2}+\alpha_{i}^{2}(\omega
_{n},\mu)\right]  \right)  ^{-s}\right\} \nonumber\\
& \nonumber\\
&  =-\lim_{s\rightarrow0}\partial_{s}\left\{  \left(  {\Lambda}\right)
^{2s}\sum\limits_{n=-\infty}^{+\infty}\frac{\Gamma(s-3/2)}{8\pi^{\frac{3}{2}%
}\Gamma(s)}\left(  \alpha_{i}^{2}(\omega_{n},\mu)\right)  ^{\frac{3}{2}%
-s}\right\}
\end{align}
so that we get
\begin{equation}
I^{(\alpha_{i})}=-\lim_{s\rightarrow0}\partial_{s}\left\{  \left(  {\Lambda
}\right)  ^{2s}\sum\limits_{n=-\infty}^{+\infty}\frac{1}{8\pi^{\frac{3}{2}%
}\Gamma(s)}\int_{0}^{\infty}{dt~e^{-t\alpha_{i}^{2}}t^{s-5/2}}\right\}  .
\end{equation}
In order to proceed, it is necessary to have an explicit form for the
$\alpha_{i}$, it can be shown that (see Appendix \ref{I_computation} for
details) $\alpha_{i}^{2}=m_{i}^{2}-(\mu+i\omega_{n})^{2},~i=(1,2,3)$. Being
$Re(m^{2})>0$ (remember the $\alpha_{i}^{2}$ are the opposite of the roots of
the numerator) the integral is convergent, thus
\begin{align}
I^{(\alpha_{i})}  &  =-\lim_{s\rightarrow0}\partial_{s}\left\{  \left(
{\Lambda}\right)  ^{2s}\sum\limits_{n=-\infty}^{+\infty}\frac{T^{3-2s}}%
{4^{s}\pi^{2s-\frac{3}{2}}\Gamma(s)}\int_{0}^{\infty}{dyy^{s-5/2}e^{-q_{i}%
^{2}y}\sum_{n=-\infty}^{\infty}e^{-y(n+c)^{2}}}\right\} \nonumber\\
&  =-\lim_{s\rightarrow0}\partial_{s}\left\{  \left(  {\Lambda}\right)
^{2s}\frac{T^{3-2s}}{4^{s}\pi^{2s-\frac{3}{2}}\Gamma(s)}\int_{0}^{\infty
}{dyy^{s-5/2}e^{-q_{i}^{2}y}\sqrt{\frac{\pi}{y}}\vartheta_{3}\left(
c\pi,e^{-\pi^{2}/y}\right)  }\right\}  ,
\end{align}
with $c=\frac{1}{2}-i\frac{\mu}{2\pi T},\ \ q_{i}^{2}=\frac{m_{i}^{2}}%
{4\pi^{2}T^{2}}$ and $y=4\pi^{2}T^{2}t$ and $\vartheta_{3}\left(  x,y\right)
$ the \textit{Jacobi $\theta$-function}. Using the well-known representation
for the Jacobi $\theta$-function
\[
\vartheta_{3}\left(  x,y\right)  =1+2\sum_{n=1}^{\infty}y^{n^{2}}\cos{2nx},
\]
the integral can be reduced to {\footnotesize
\begin{align}
I^{(\alpha_{i})}  &  =-\lim_{s\rightarrow0}\partial_{s}\left\{  \left(
{\Lambda}\right)  ^{2s}\left[  \frac{T^{3-2s}(q_{i}^{2})^{2-s}\Gamma
(s-2)}{4^{s}\pi^{2s-{2}}\Gamma(s)}+\frac{T^{3-2s}(q_{i}^{2})^{1-\frac{s}{2}}%
}{4^{s-1}\pi^{s}\Gamma(s)}\sum\limits_{n=1}^{+\infty}n^{s-2}K_{2-s}(2n\pi
\sqrt{q^{2}})\cos{\left(  n\pi-i\frac{n\mu}{T}\right)  }\right]  \right\}
\nonumber\label{I_function}\\
&  =\frac{(m_{i}^{2})^{2}}{32\pi^{2}T}\left( \log{\left( \frac{m_{i}^{2}%
}{\Lambda^{2}}\right) }-\frac{3}{2}\right)  + \sum\limits_{n=1}^{+\infty}%
\frac{m_{i}^{2}(-1)^{n}T}{\pi^{2}n^{2}}K_{2}\left(  n\frac{\sqrt{m_{i}^{2}}%
}{T}\right)  \cosh\left(  \frac{\mu n}{T}\right) ,
\end{align}
} where we assume $\mu<Re(m_{i})$.

\subsection{Existence of equation of state}

From the partition function, we can obtain the pressure, entropy density,
number density and energy density, respectively:
\begin{align}
P(T,\mu)  &  =\frac{T}{V}\log Z(T,\mu),\nonumber\label{thermodynamics}\\
s(T,\mu)  &  =\frac{\partial P}{\partial T}(T,\mu),\\
n(T,\mu)  &  =\frac{\partial P}{\partial\mu}(T,\mu),\nonumber\\
e(T,\mu)  &  =Ts-P+\mu n.\nonumber
\end{align}
Inverting the last equation, we can obtain the EOS either for $\mu$ constant
$P=P(e,T)\big{|}_{\mu}$ or $T$ constant $P=P(e,\mu)\big{|}_{T}$. We stress
here the fact that, if the pressure and energy density function in
\eqref{thermodynamics} are not strictly monotonic functions in some interval
$J\in\mathbb{R}$ either for $T$ or $\mu$, and also the shape of
non-monotonicity in $J$ does not coincide exactly for $P$ and $e$, then it is
not possible to define a proper EOS as a function $P=P(e)$ in the interval $J$.

\begin{figure}[ptb]
\begin{center}
\includegraphics[width=.6\textwidth]{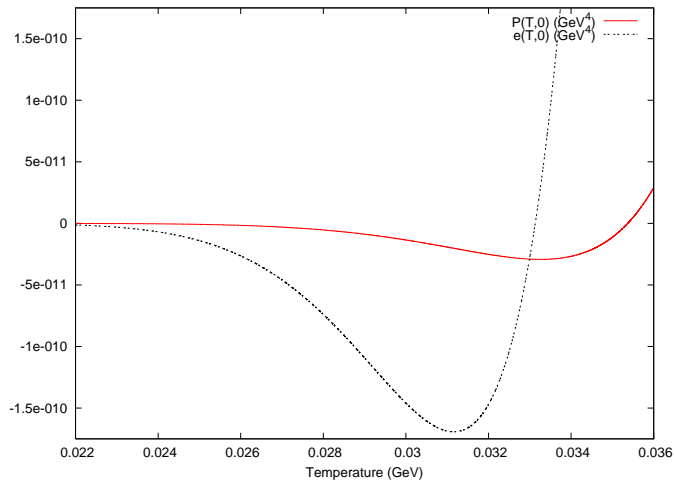}
\end{center}
\caption{The pressure $P$ of the quark sector (red line) and its energy
density $e$ (black dots) as function of the temperature for $\mu=0$. It is
worth to note the detail of the plot due the critical zone is very narrow
compared to the entire unit range and the values of negative $P$ is less than
$2\times10^{-5}$ GeV.}%
\label{pres_ener_vs_temp_norm}%
\end{figure}

\begin{figure}[ptb]
\begin{center}
\includegraphics[width=.6\textwidth]{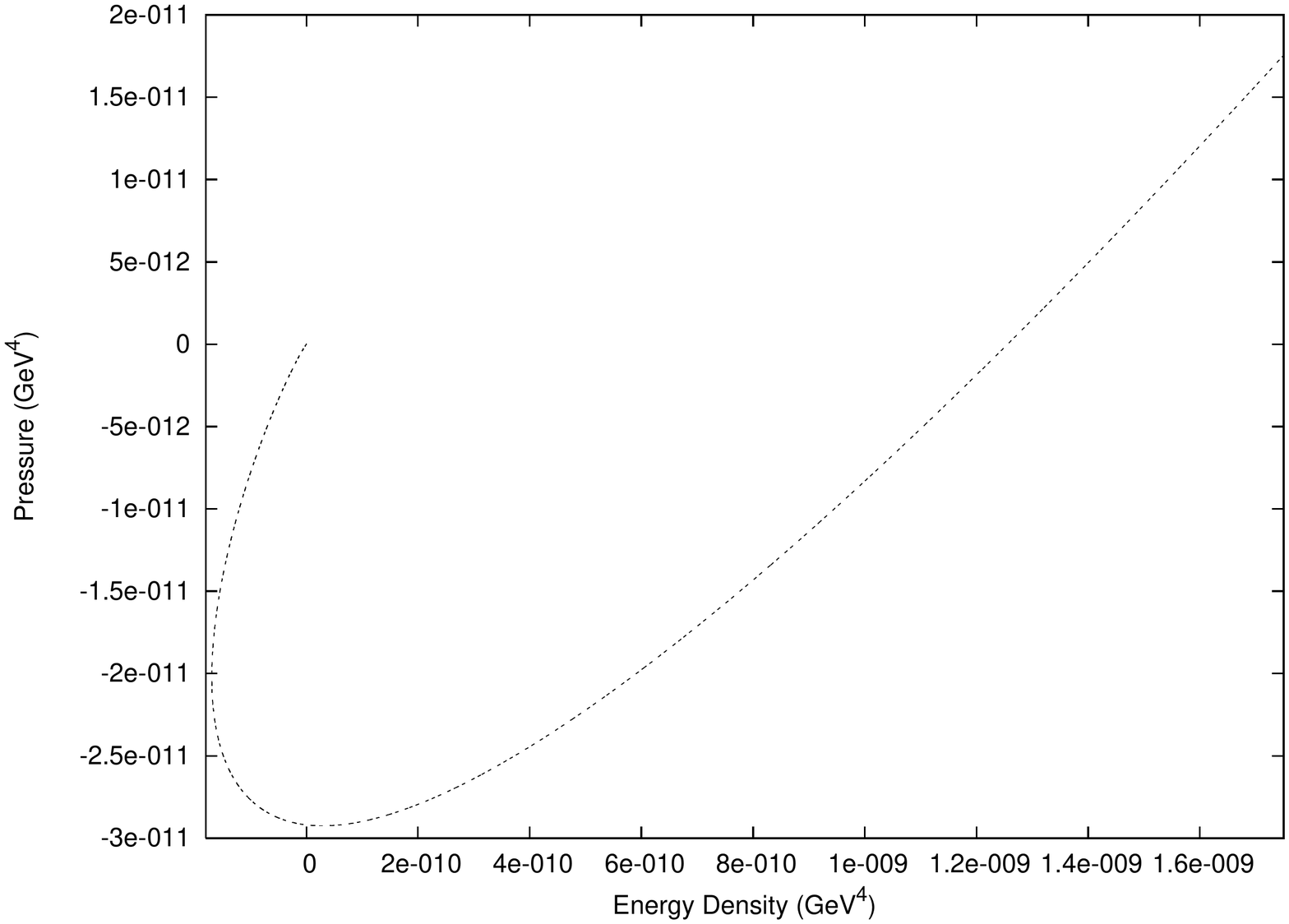}
\end{center}
\caption{The pressure $P$ of the quark sector as function of the energy
density $e$ for $\mu=0$. We can see if $e\leq0$ then the EOS cannot be
defined.}%
\label{pres_vs_ener_crit}%
\end{figure}

\begin{figure}[ptb]
\begin{center}
\includegraphics[width=.5\textwidth]{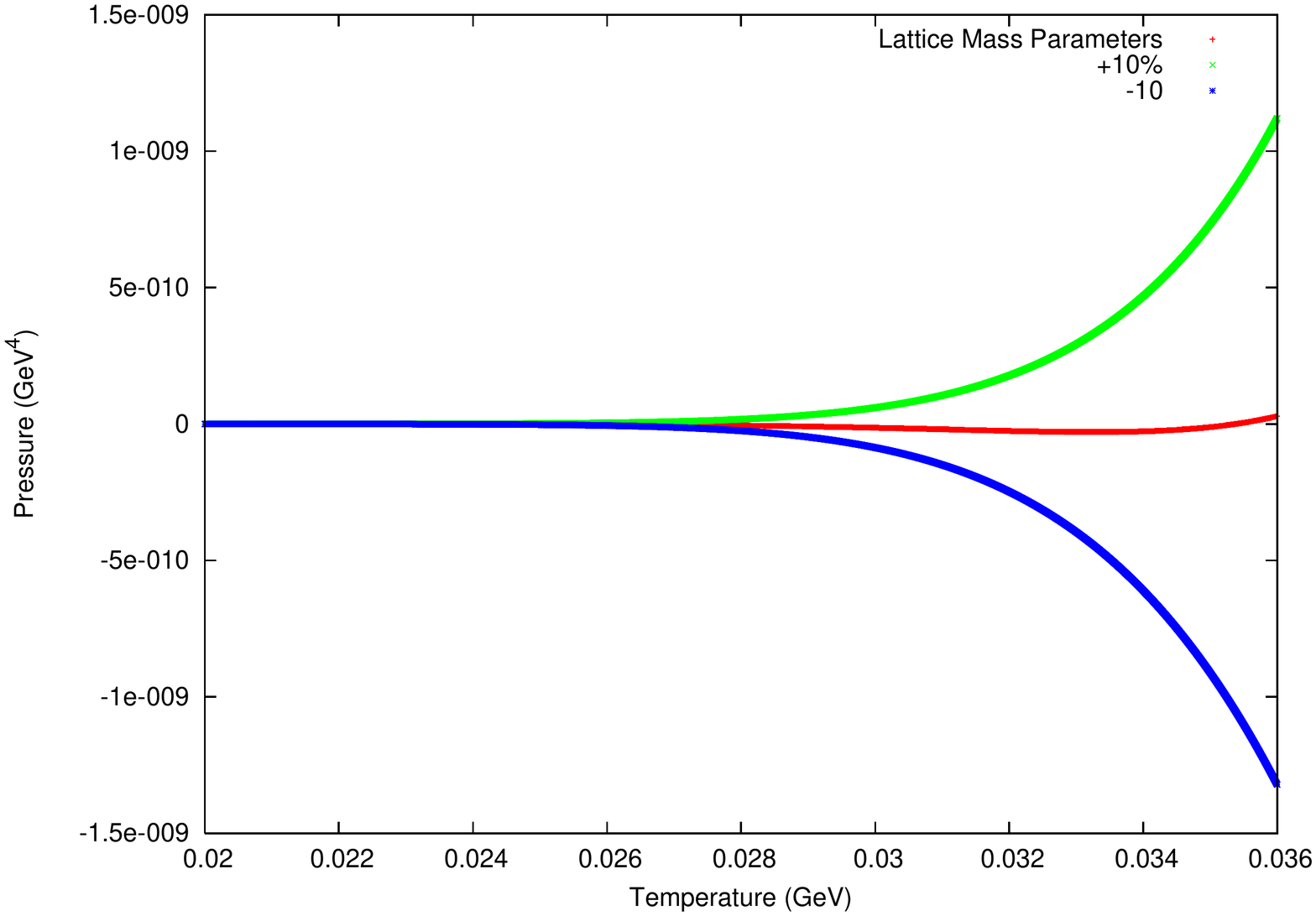}
\end{center}
\caption{Comparison of the pressure of the quark sector as function of
temperature in the critical zone for $\mu=0$, $M_{3}$, $m^{2}$ and $m_{0}$
given in \ref{quarkp2}, and a $\pm10\%$ modification of these lattice mass
parameters.}%
\label{pres_vs_temp_norm_modif}%
\end{figure}

As is shown in Figure \ref{pres_ener_vs_temp_norm}, using the mass fit
parameters \eqref{quarkp2} for the quark propagator, we see there is a range
of $T\in(0.0,3.36)\times10^{-2}~ \hbox{GeV}$, which we will call \emph{critical
zone}, where we cannot find an EOS. This is because when we have imaginary a
component of the poles of the propagator, the modified Bessel functions
$K_{\nu}$ acquire, below $T=0.036~\hbox{GeV}$, an oscillatory behavior, see Appendix
\ref{considerations}. In our opinion, these results suggest that, at low
temperature, some extra physical parameters (which, for example, can be
related to light glueballs) suitable to properly label the equilibrium states are
needed (see Appendix \ref{considerations} for some considerations on this
respect). For instance, there is not a unique value of the normalized pressure
for a normalized energy density of $-5\times10^{-11}$. 
It is interesting to address the plot of pressure $P$ as a function of the
energy density $e$. As is shown in Figure \ref{pres_vs_ener_crit}, there is
an EOS when both $e$ and $P$ are positive. 
On the other hand, if we change the fit parameters $+10~\%$, such
critical zone would disappear (or it could even increase if we decrease the
lattice parameters by $-10~\%$), as we can compare in Figure
\ref{pres_vs_temp_norm_modif}.

\subsection{Checking causality and Le Chatelier's principles}

\begin{figure}[ptb]
\begin{center}
\includegraphics[width=.5\textwidth]{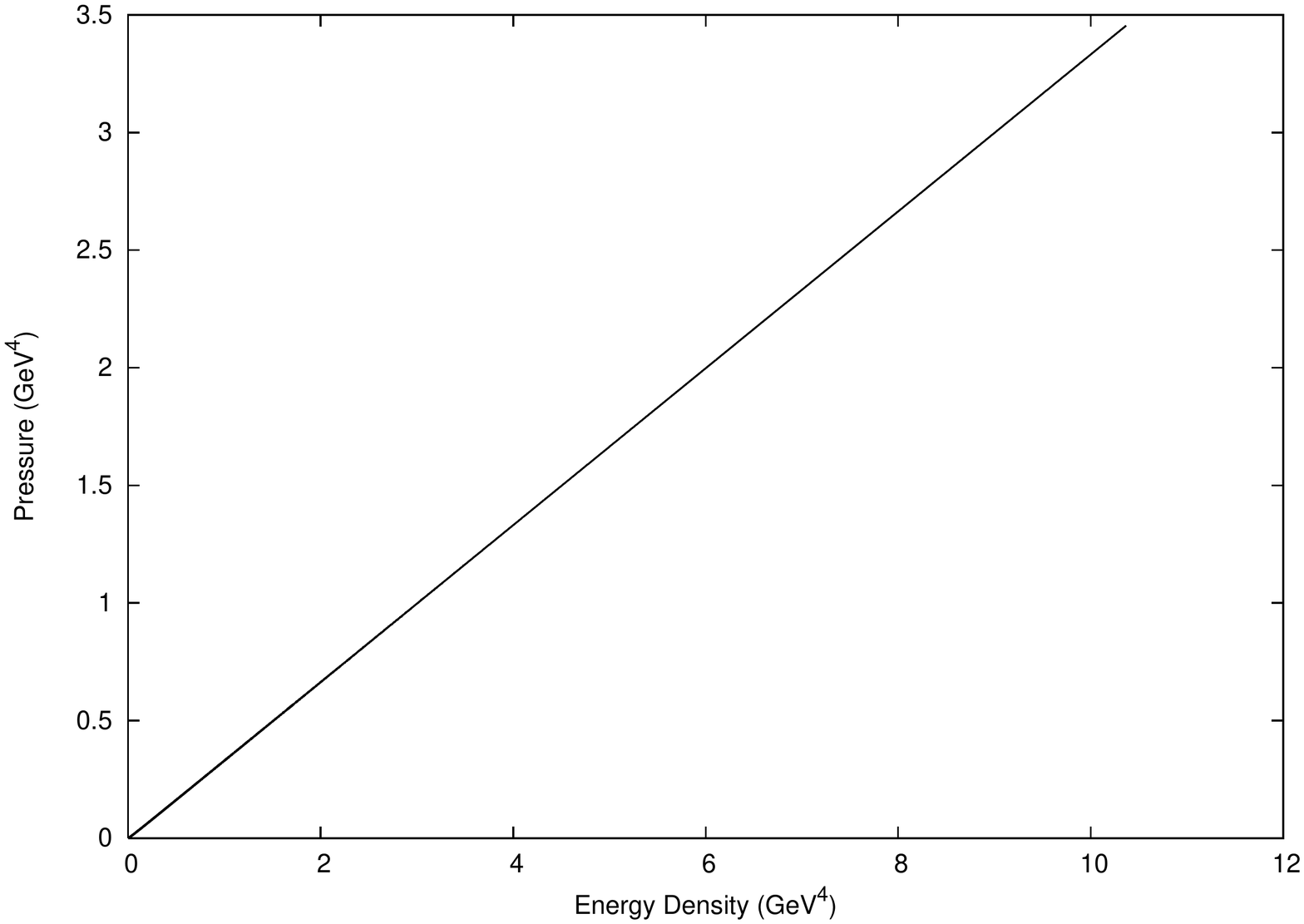}
\end{center}
\caption{The pressure $P$ of the quark sector as a function of the energy $e$
for $\mu=0$.}%
\label{pres_vs_energy}%
\end{figure}

\begin{figure}[ptb]
\begin{center}
\includegraphics[width=.5\textwidth]{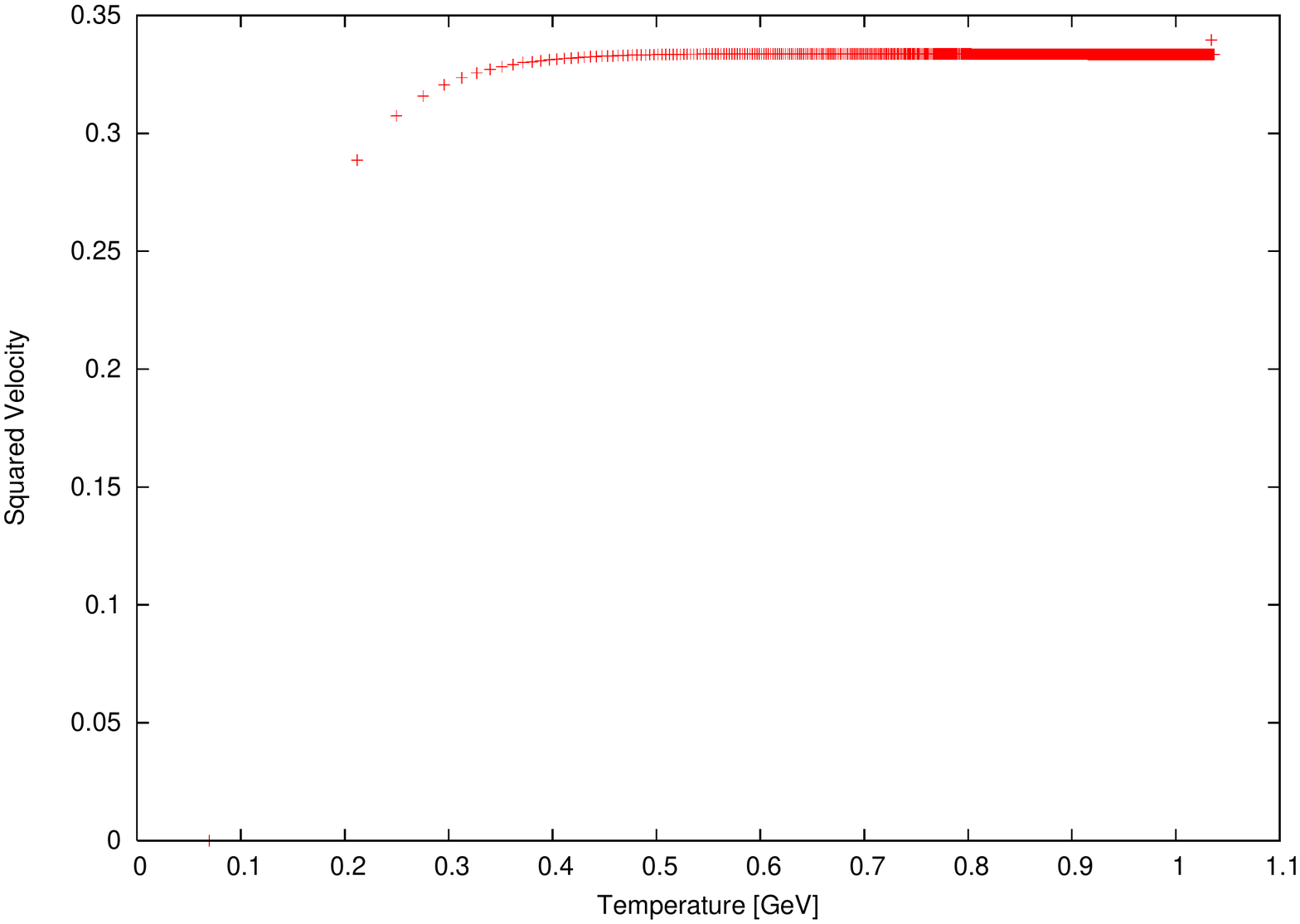}
\end{center}
\caption{The squared sound velocity of the quark sector as function of the
temperature for $\mu=0$.}%
\label{veloc_vs_temp}%
\end{figure}

Once we can write the EOS, we may wonder if the conditions of causality
($c^{2}=\frac{\partial P}{\partial e}\leq1$) and Le Chatelier's principle ($P$
is monotonically growing with respect to $e$), which appearing in \cite{ruffini},
are both satisfied. As we can see in Figure \ref{pres_vs_energy}, Le
Chatelier's principle holds for the mass values of the lattice data, when it
is possible to define an EOS $P=P(e)$ (see description above).
In Figure \ref{veloc_vs_temp}, one can see that causality holds as
well. 

\section{The gluonic sector}

In this section we will check, as we did in the previous section for the
quarks, whether or not the non-perturbative gluons propagator in the Landau
gauge arising from the lattice \cite{cucchieri1} gives rise to an EOS
satisfying the three principle mentioned above. The extra benefit of this
analysis is that, in the gluons case, the non-perturbative propagator arising
from the lattice data can also be deduced theoretically. Such a propagator is
strongly related with one of the most fascinating non-perturbative effects in
non-Abelian gauge theories: the appearance of Gribov copies
\cite{Gribov:1977wm}. On flat, topologically trivial, space-times (which is
the only case which will be analyzed here) Gribov copies represent a
topological obstruction to define globally the gauge-fixing peculiar of the
non-Abelian nature of the gauge group (for a good review see \cite{SS05}).

On the other hand, both on curved spaces and on flat spaces but with
non-trivial topology the pattern of appearance of Gribov copies can be
considerably more complicated. For instance, Gribov copies can appear even in
the Abelian case (see in particular
\cite{nostro1,nostro2,nostro3,nostro4,decesare,nostro5,nostro5.5}). The
results in these references suggest, as a future direction, to
analyze how the equations of state of interacting gauge bosons (\textit{even
in the Abelian case}) depend on the non-trivial backgrounds on which they are
defined. We hope to come back on this interesting issue in a future publication.

The most effective method to eliminate Gribov copies (proposed by Gribov
himself in \cite{Gribov:1977wm} and refined in
\cite{Zwanziger:1988jt,Zwanziger:1989mf,Zwanziger:1992qr}) amounts to restrict
the domain of the path integral to the region in which the Faddeev-Popov
operator is positive definite (such region is called Gribov region). In
\cite{DZ89} the authors showed that all the gauge orbits cross the Gribov
region. Hence, the GZ restriction does not lose any relevant physical
configuration. A local and renormalizable effective action for Yang-Mills
theory whose dynamics is restricted to the Gribov horizon was constructed in
\cite{local1, local2,local3,local4} by adding extra fields to the action.
Later, an improved action was proposed by considering suitable condensates
\cite{vandersickel}, which leads to propagators and glueball masses in
agreement with the lattice data \cite{sorellaPRL}. With the same action, one
can also solve the old problem of the Casimir energy in the MIT-bag model
\cite{nostro6}. Moreover, this approach is quite effective also at finite
temperature \cite{zwat1,nostro7,zwat2} and, at least at one-loop order, gives
rise to a vacuum expectation value for the Polyakov loop compatible with its
role of order parameter for the confinement-deconfinement transition \cite{polyakov-paper}.

\subsection{Gribov-Zwanziger approach}

Within the GZ approach, the vacuum energy density at one loop can be written
as (we will follow the notation of \cite{polyakov-paper}):
\begin{equation}
\mathcal{E}_{v}=-\frac{d(N^{2}-1)}{2Ng^{2}}\lambda^{4}+\frac{1}{2\beta
V}(d-1)\ln\left( \frac{D^{4}+\lambda^{4}}{\Lambda^{4}}\right) -\frac{d}{2\beta
V}\ln\left( \frac{-D^{2}}{\Lambda^{2}}\right) \;, \label{vacuum-gz}%
\end{equation}
being $\Lambda$ a regularization parameter, $\beta=1/T$ and $V$ the spatial
volume. One can normalize the Gribov parameter $\lambda$ in order to absorb
the divergent part of \eqref{vacuum-gz}. In this way one obtains for $SU(2)$
internal gauge group
\begin{equation}
\mathcal{E}_{v}(T)=\frac{3}{2}(d-1)\left[  I(T,i\lambda^{2})+I(T,-i\lambda
^{2})\right] -\frac{d}{2}I(T,0) , \label{vacuum-gz-renormalized}%
\end{equation}
where in the $d=4$ case
\begin{align}
\label{I_function_gluon}I(T,\alpha^{2}) & =T\sum\limits_{n=-\infty}^{+\infty
}\int\frac{dp^{3}}{(2\pi)^{3}}\ln\Lambda^{2}\left( \omega_{n}^{2}+m^{2}%
+\vec{p}^{2}\right) \nonumber\\
& =\frac{(\alpha^{2})^{2}}{32\pi^{2}}\left(  \ln\left(  \frac{\alpha^{2}%
}{\Lambda^{2}}\right)  -\frac{3}{2}\right)  -\frac{\alpha^{2}T^{2}}{\pi^{2}%
}\sum\limits_{n=1}^{+\infty}(-1)n^{-2}K_{2}(n\frac{m}{T}),
\end{align}
where we have taken the thermodynamic limit $V\to+\infty$ in the second
equality, which implies $\sum\limits_{q}\to V\int\frac{d^{3}q}{(2\pi)^{3}}$
\cite{SS05}. The detailed computations of \eqref{I_function_gluon} are in
Appendix \ref{I_computation}.

\begin{figure}[ptb]
\begin{center}
\includegraphics[width=.5\textwidth]{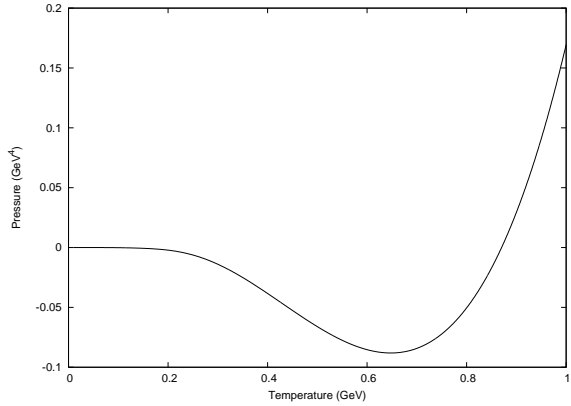}
\end{center}
\caption{The pressure $P$ as a function of the temperature $T$ in GZ approach.
It can be seen a range of $T$ where $P$ is negative.}%
\label{pressure_vs_temp_GZ}%
\end{figure}

\begin{figure}[ptb]
\subfloat[][]{\includegraphics[width=.5\textwidth]{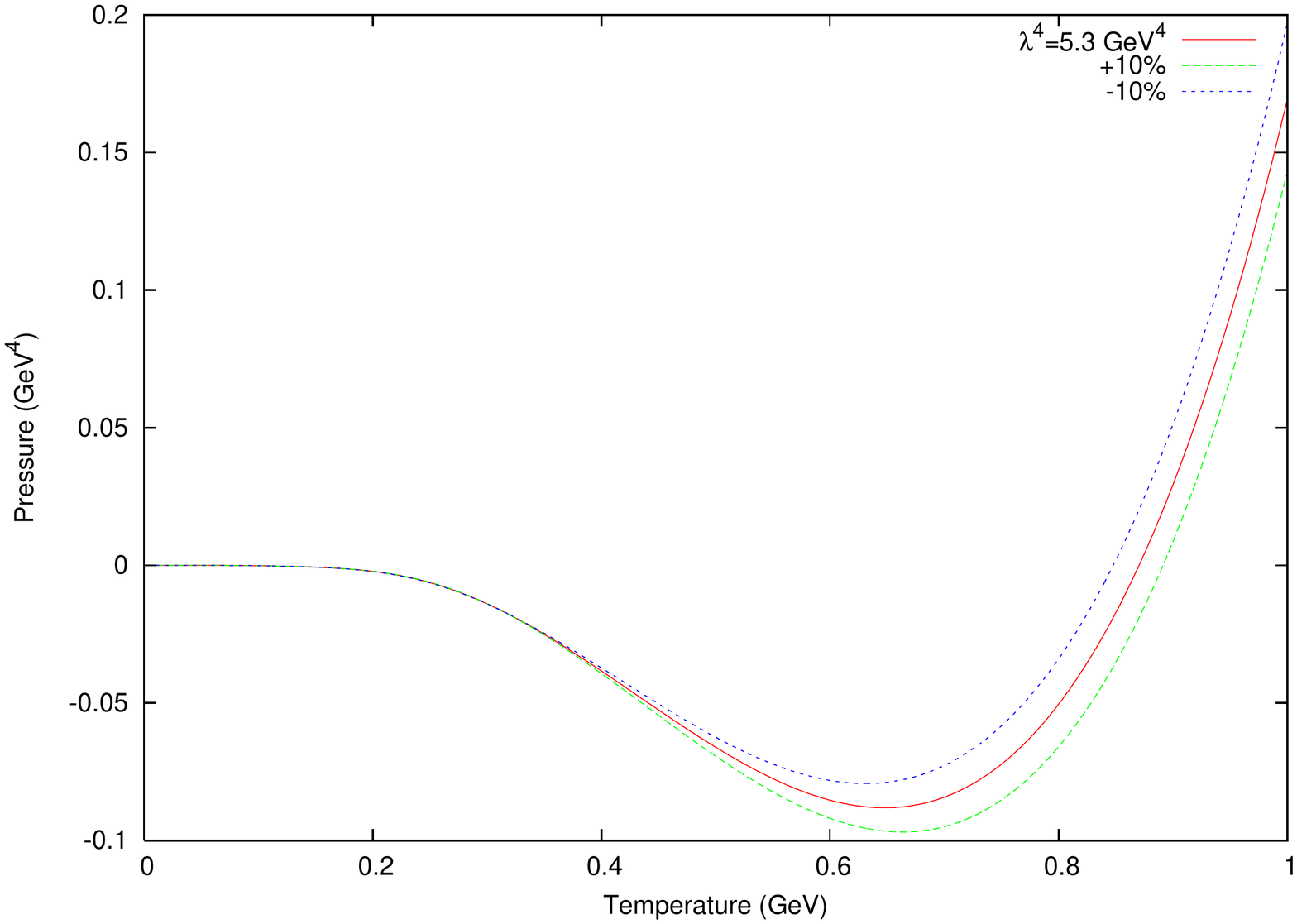}}
\subfloat[][]{\includegraphics[width=.5\textwidth]{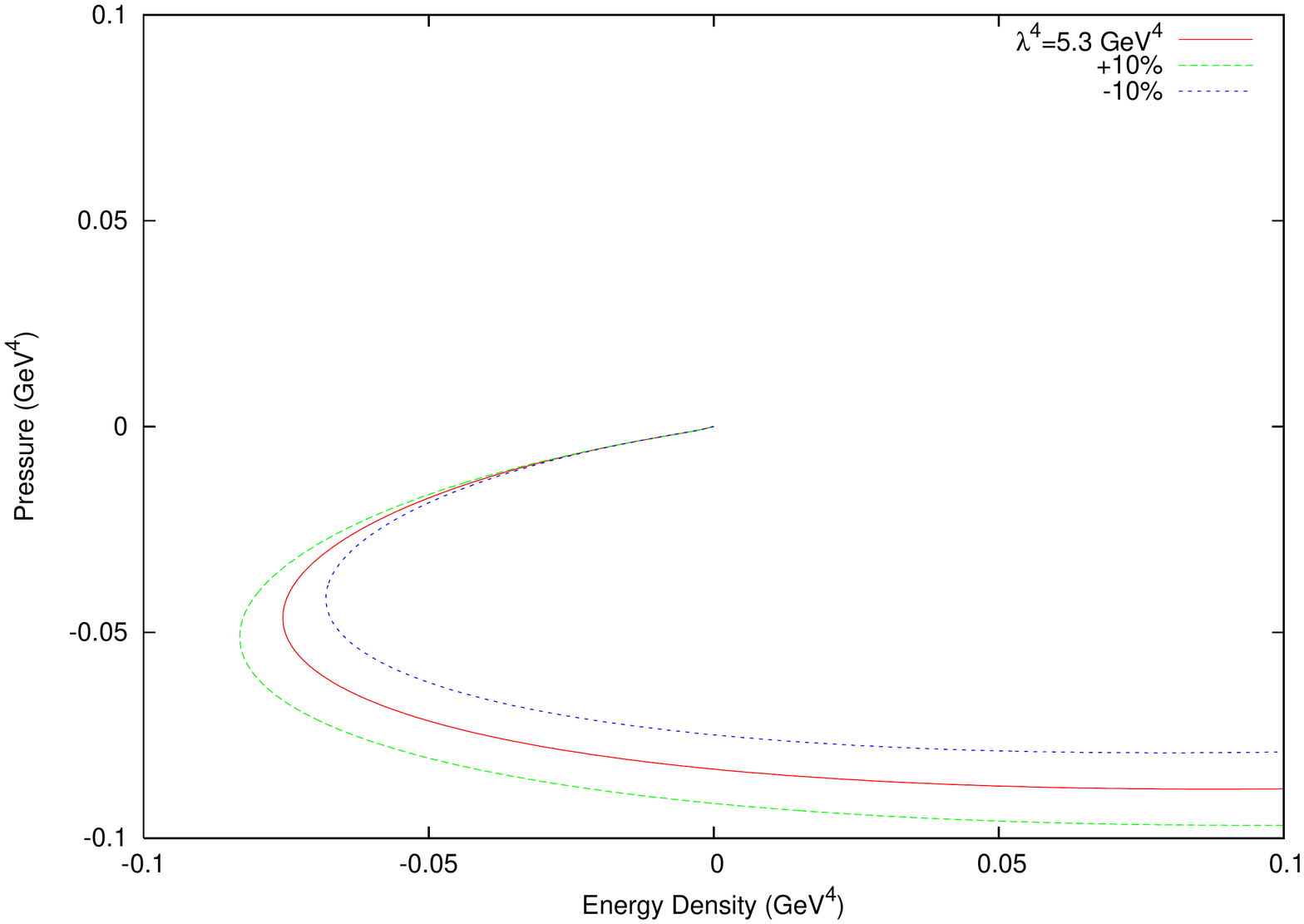}}
\caption{(a) The gluon pressure as a function on temperature for different
values of the Gribov parameter $\lambda$ in GZ approach. (b) Pressure as a
function of energy density for different values of the Gribov parameter
$\lambda$ in GZ approach.}%
\label{pres_vs_temp_modif_GZ}%
\end{figure}

\begin{figure}[ptb]
\begin{center}
\includegraphics[width=.5\textwidth]{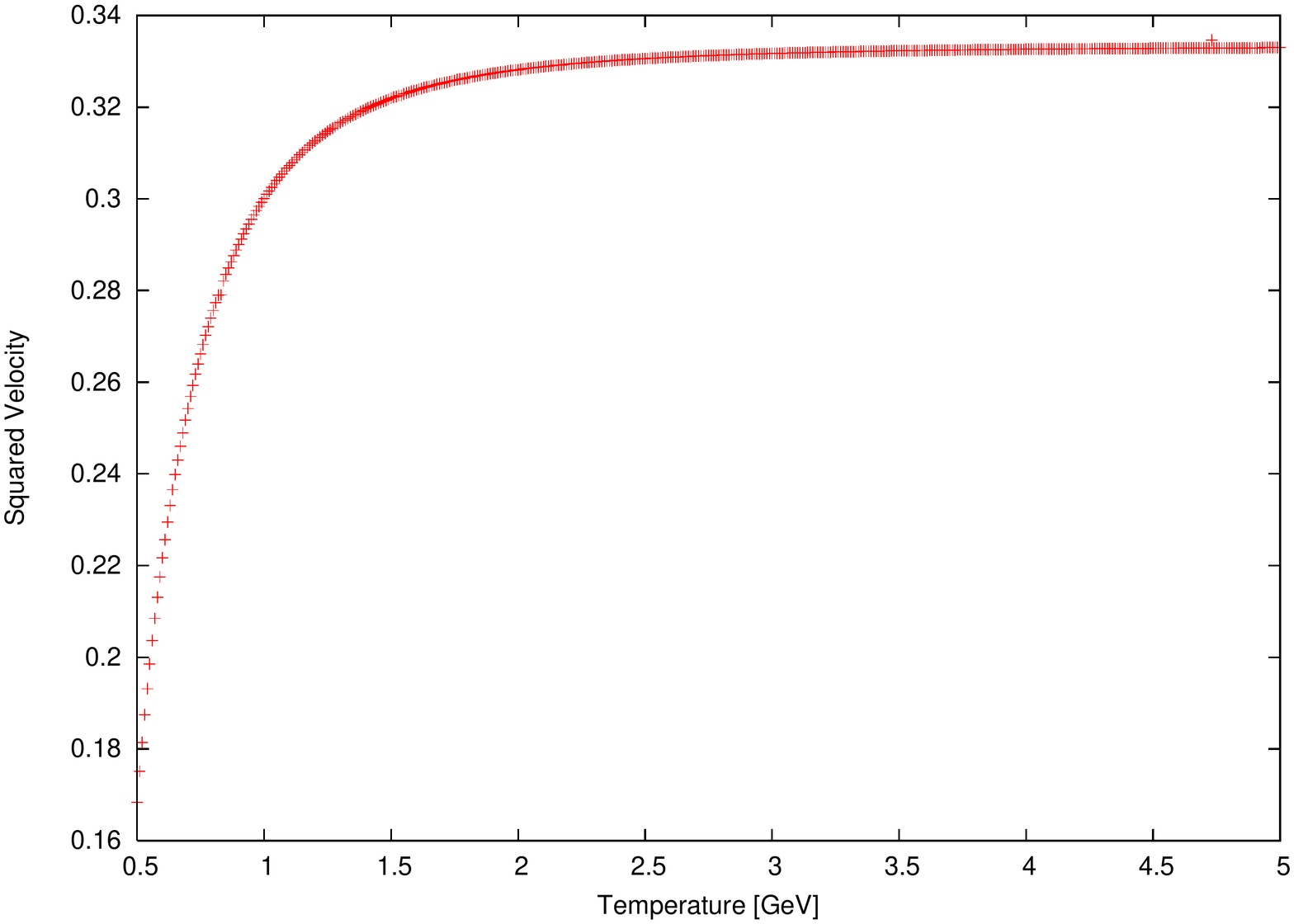}
\end{center}
\caption{The squared velocity as a function of the temperature in GZ
approach.}%
\label{vel_vs_temp_GZ}%
\end{figure}

\begin{figure}[ptb]
\subfloat[][]{\includegraphics[width=.5\textwidth]{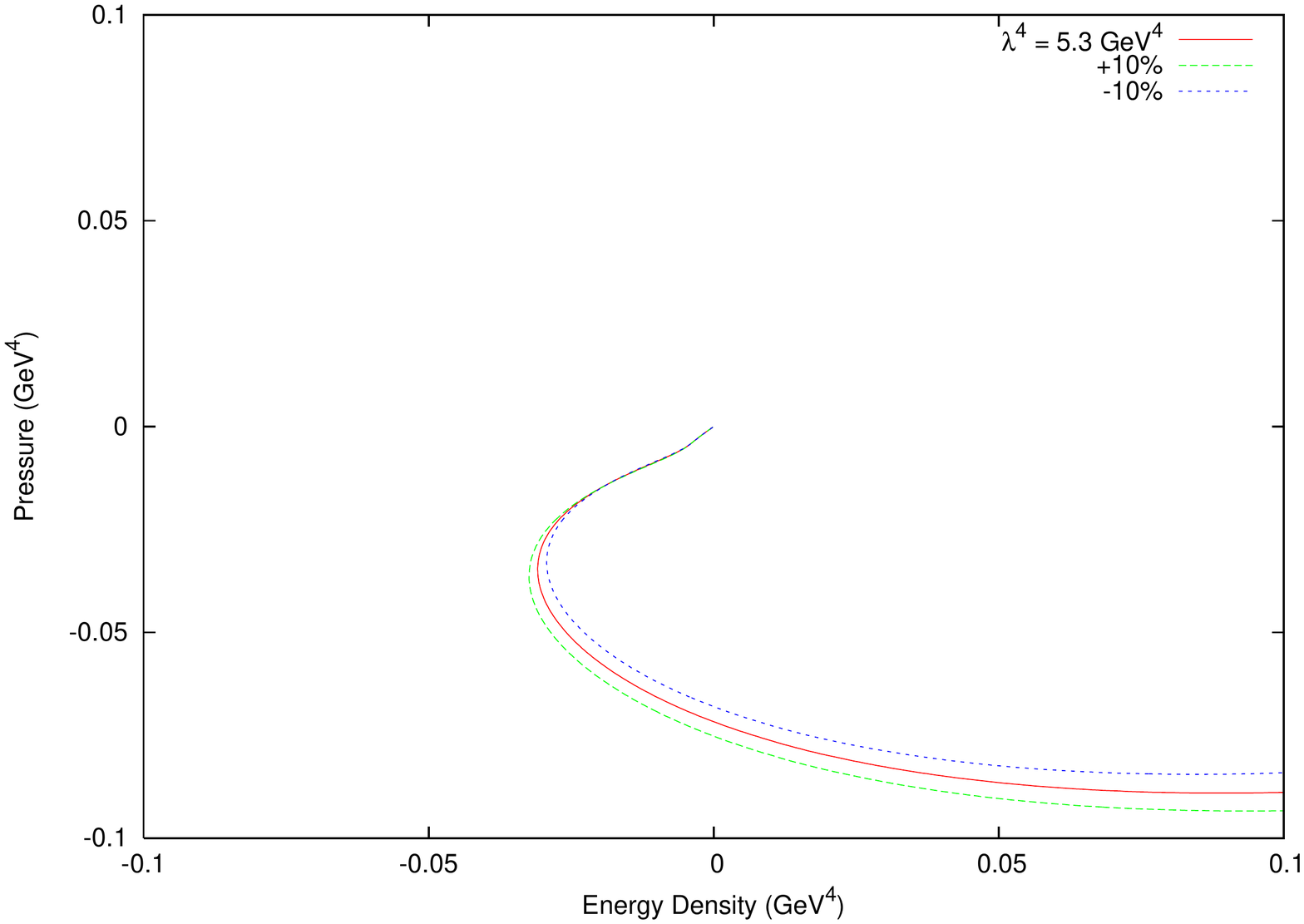}}
\subfloat[][]{\includegraphics[width=.5\textwidth]{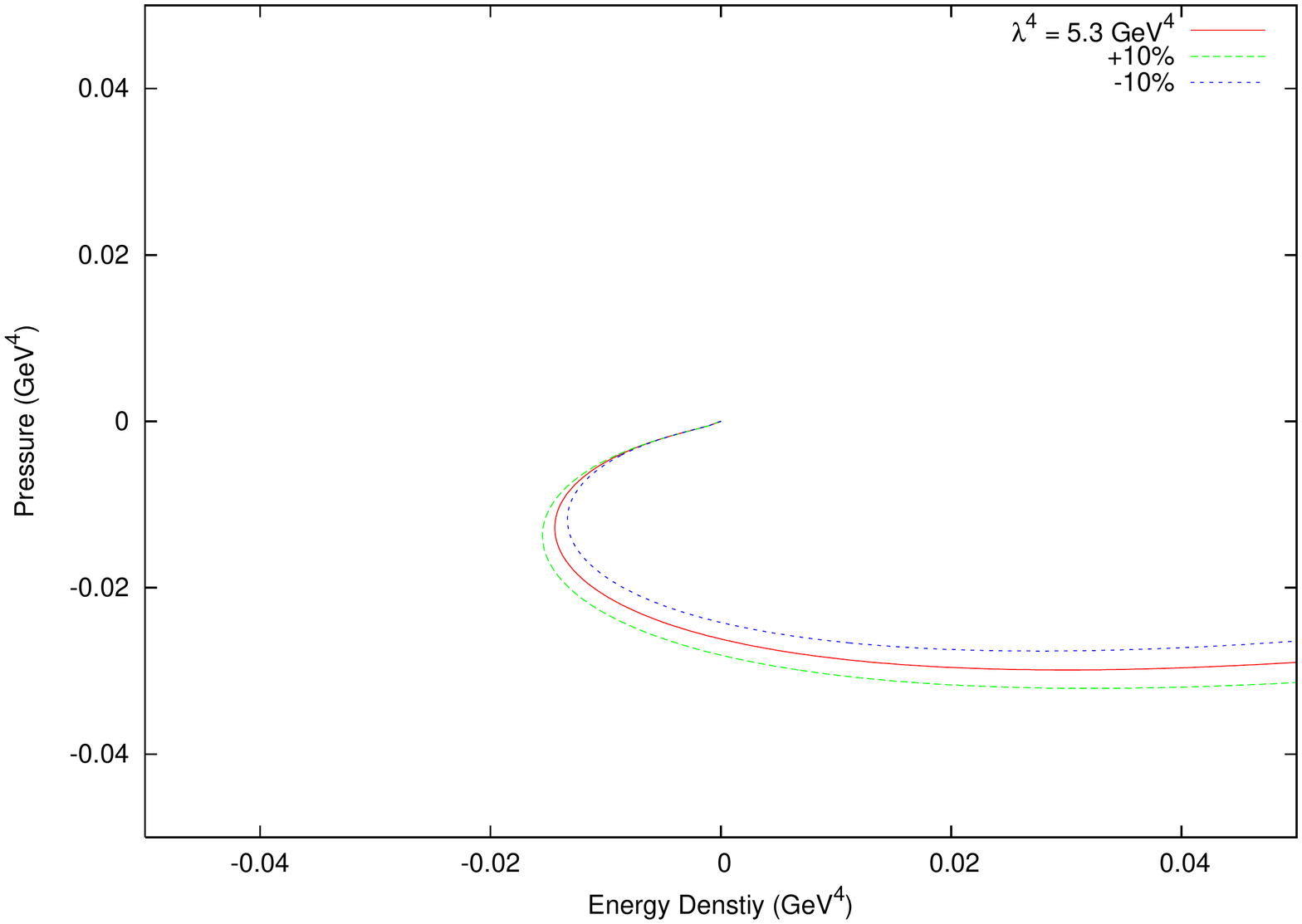}}\caption{(a)
Pressure as a function of energy density and different values of the Gribov
parameter $\lambda$ in GZ approach in dimension $d=2$. We take the
$\lambda^{4}$ value from \cite{Cucchieri:2011ig}. (b) Pressure as a function
of energy density and different values of the Gribov parameter $\lambda$ in GZ
approach in dimension $d=3$. We take the $\lambda^{4}$ value from
\cite{Cucchieri:2011ig}.}%
\label{pres_vs_ener_GZ_d2_d3}%
\end{figure}

In principle, as we are working at finite temperature, one should first
determine how the Gribov mass parameter $\lambda$ depends on the temperature
itself (see \cite{zwat1,nostro7,zwat2}). However, this would lead to a coupled
system of integral equations to be solved self-consistently and this would
enormously complicate the numerical task. Fortunately, as the available
results clearly indicate, the Gribov parameter $\lambda$ changes very slowly
in the range of temperatures analyzed in the present paper (see Figure 2 of
\cite{polyakov-paper} where $\lambda$ was computed as a function of $T$, the
only reasonable assumption being that the coupling constant $g$ does not
change in a small range around $T=0$) so that we will work in the
approximation in which $\lambda$ does not depend on the temperature and is
actually equal to its $T=0$ value (see \cite{Cucchieri:2011ig}, where
$\lambda^{4}=5.3~\hbox{GeV}^{4}$). In such a case, we can construct
$P(T)=-{\mathcal{E}}_{v}(T)$ and the result is plotted in Figure
\ref{pressure_vs_temp_GZ}. 
Interestingly, as was also suggested by the results on the
Polyakov loop \cite{polyakov-paper}, there is a range of temperatures where
the pressure is negative and is not a strictly monotonic function.
Furthermore, if we modify the Gribov parameter by $\pm$ the $10~\%$ of its
value, the plot remains almost the same, as we can see in Figure
\ref{pres_vs_temp_modif_GZ}. Namely, within the GZ approach, the zeroth
consistency condition of \cite{ruffini} is likely to be always violated.
In Figure \ref{vel_vs_temp_GZ} we plotted the squared velocity as
a function of temperature in the region where we can define an EOS. We see Le
Chatelier's principle and causality conditions are both satisfied.
It is interesting to show if the zeroth consistency condition is fulfilled in
dimensions $d=2$ and $d=3$ also with $SU(2)$ internal gauge group. The
computation could be done using \eqref{vacuum-gz-renormalized} and
considering\footnote{We take only the modes $n\neq0$ for $I(T,\alpha^{2})$ in
dimensions $d=2$ and $d=3$.} for $d=2$
\begin{equation}
I(T,\alpha^{2})=-\frac{2}{\pi}T\sqrt{\alpha^{2}}\sum\limits_{n=1}^{+\infty
}n^{-1}K_{1}(n\frac{\sqrt{\alpha^{2}}}{T}),
\end{equation}
while for $d=3$
\begin{equation}
I(T,\alpha^{2})=-\frac{\sqrt{2}}{\pi^{3/2}}T^{3/2}(\alpha^{2})^{3/4}%
\sum\limits_{n=1}^{+\infty}n^{-3/3}K_{3/2}(n\frac{\sqrt{\alpha^{2}}}{T}),
\end{equation}
and using for both cases the Gribov parameter $\lambda^{4}$ given in
\cite{Cucchieri:2011ig}. As is shown in Figure \ref{pres_vs_ener_GZ_d2_d3},
both in $d=2$ and $d=3$, there is a region where the EOS cannot be well
defined in the GZ approach. 

\subsection{Refined Gribov-Zwanziger approach}

\begin{figure}[ptb]
\begin{center}
\includegraphics[width=.5\textwidth]{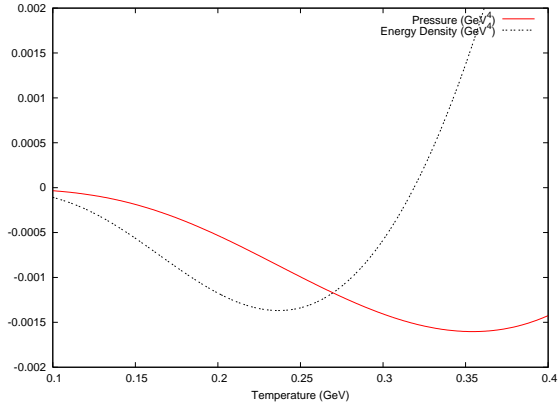}
\end{center}
\caption{The gluon pressure (red line) and gluon energy density (black dots)
as function of temperature for $\mu=0$ in RGZ approach.}%
\label{pres_energy_temp_RGZ}%
\end{figure}

\begin{figure}[ptb]
\subfloat[][]{\includegraphics[width=.5\textwidth]{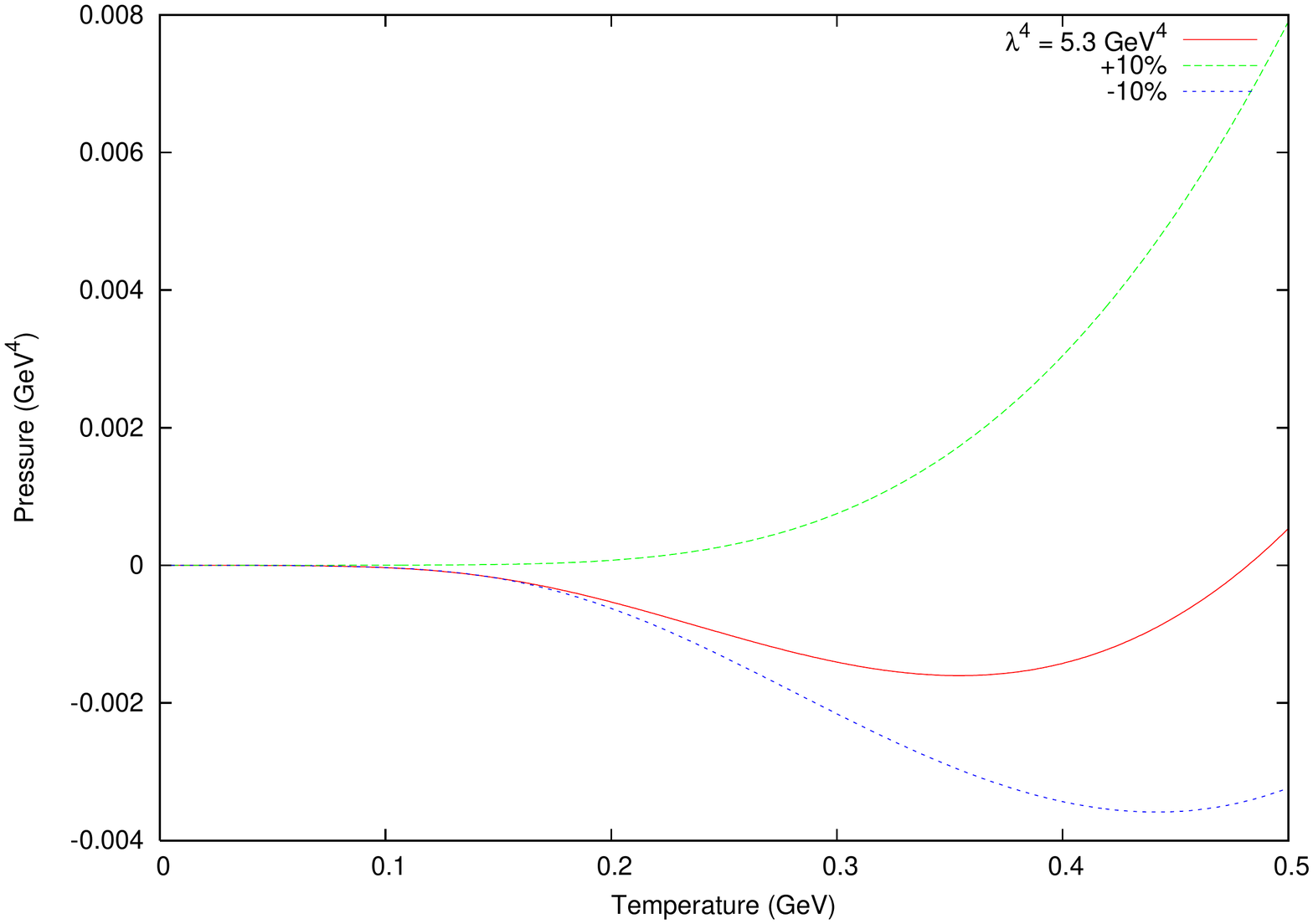}}
\subfloat[][]{\includegraphics[width=.5\textwidth]{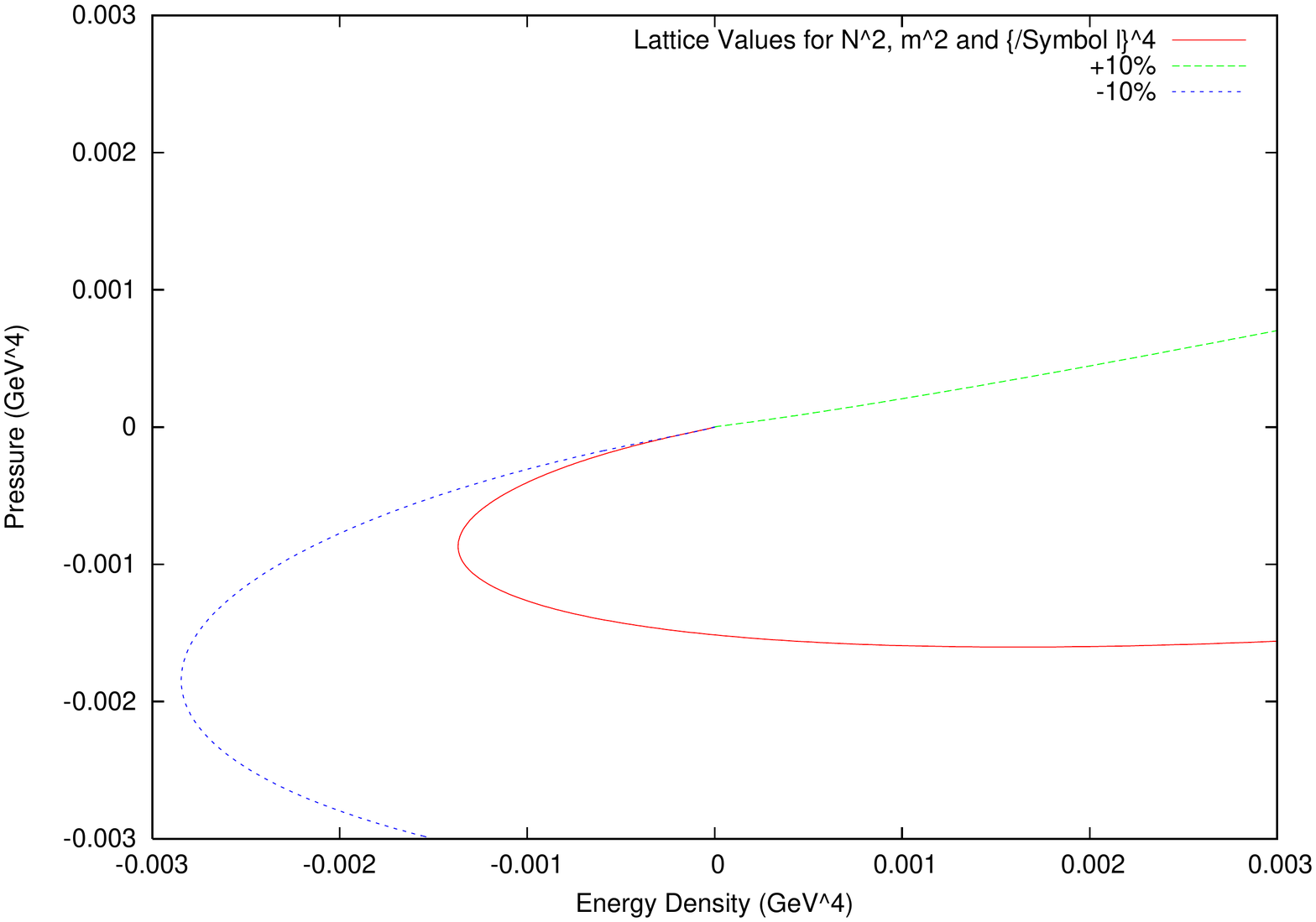}}
\caption{(a) The pressure as function on temperature for $\mu=0$ and different
values of the lattice mass parameters in RGZ approach. We see in this critical
region, the EOS is not well-defined. (b) The pressure as a function of the
energy density for different values of the RGZ parameters $N^{2}$, $m^{2}$ and
$\lambda^{4}$. We observe an EOS seems to be almost well-defined when the RGZ
parameters are modified $+10\%$.}%
\label{pres_ener_modif_RGZ}%
\end{figure}

\begin{figure}[ptb]
\subfloat[][]{\includegraphics[width=.35\textwidth]{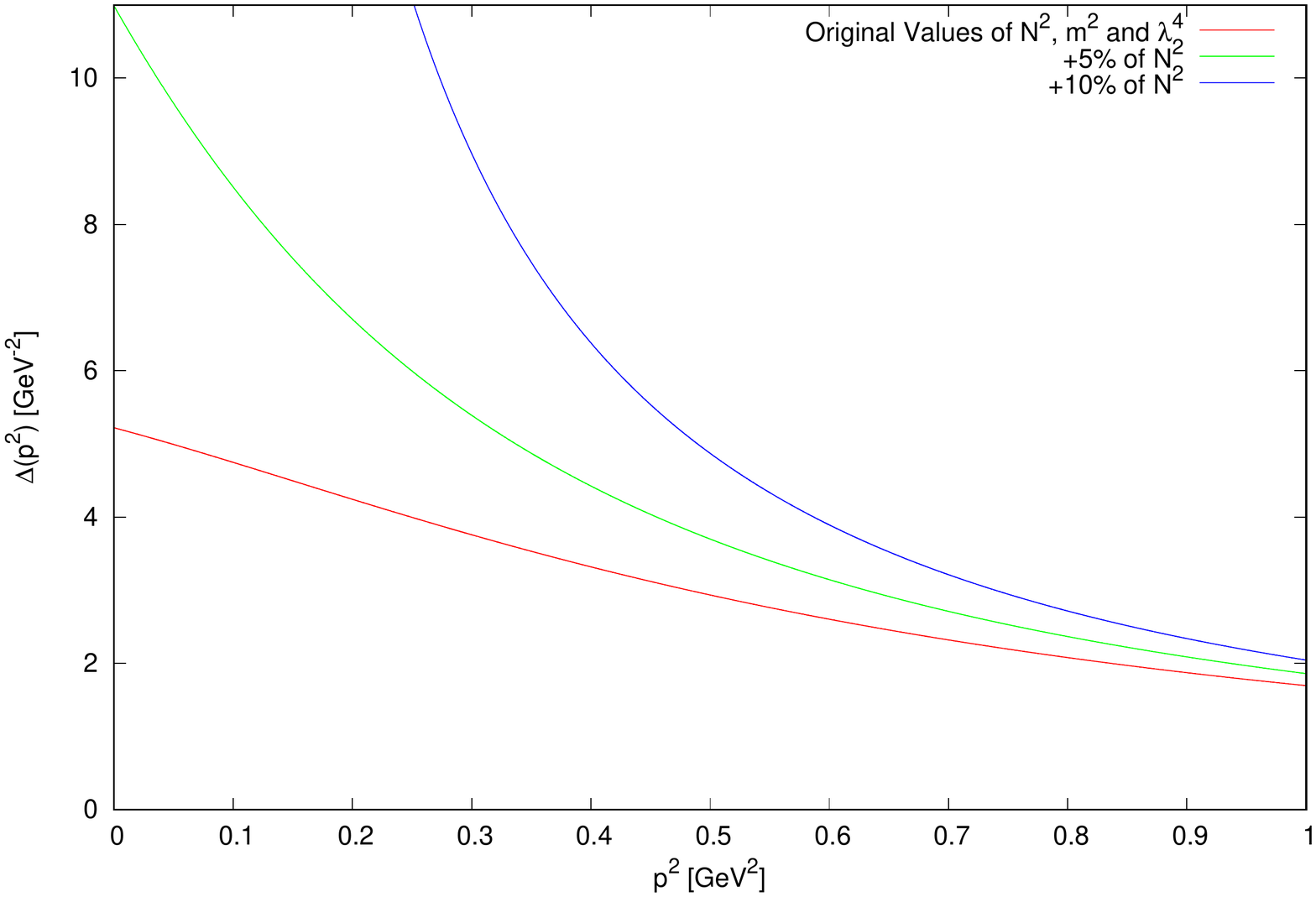}}
\subfloat[][]{\includegraphics[width=.35\textwidth]{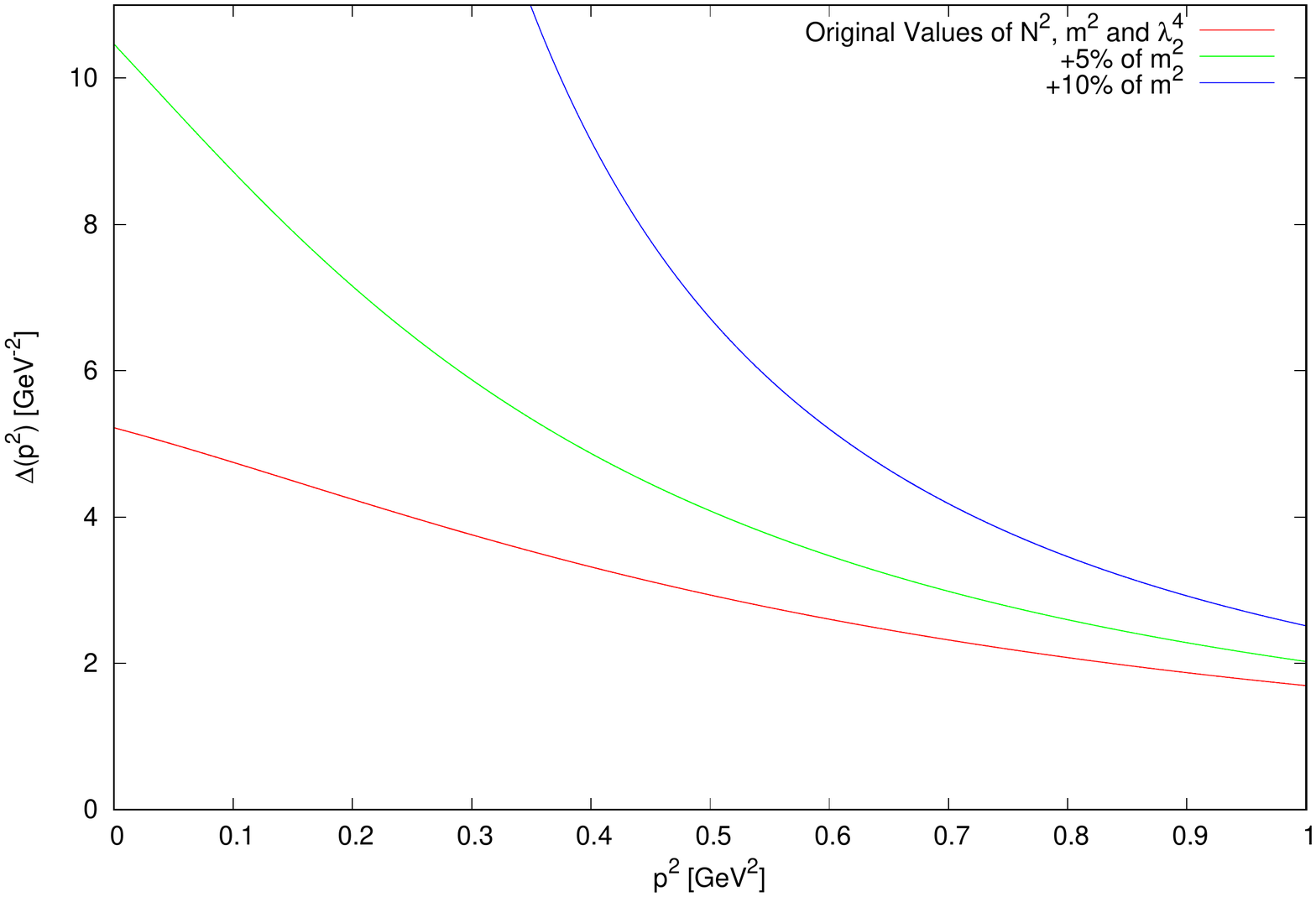}}
\subfloat[][]{\includegraphics[width=.35\textwidth]{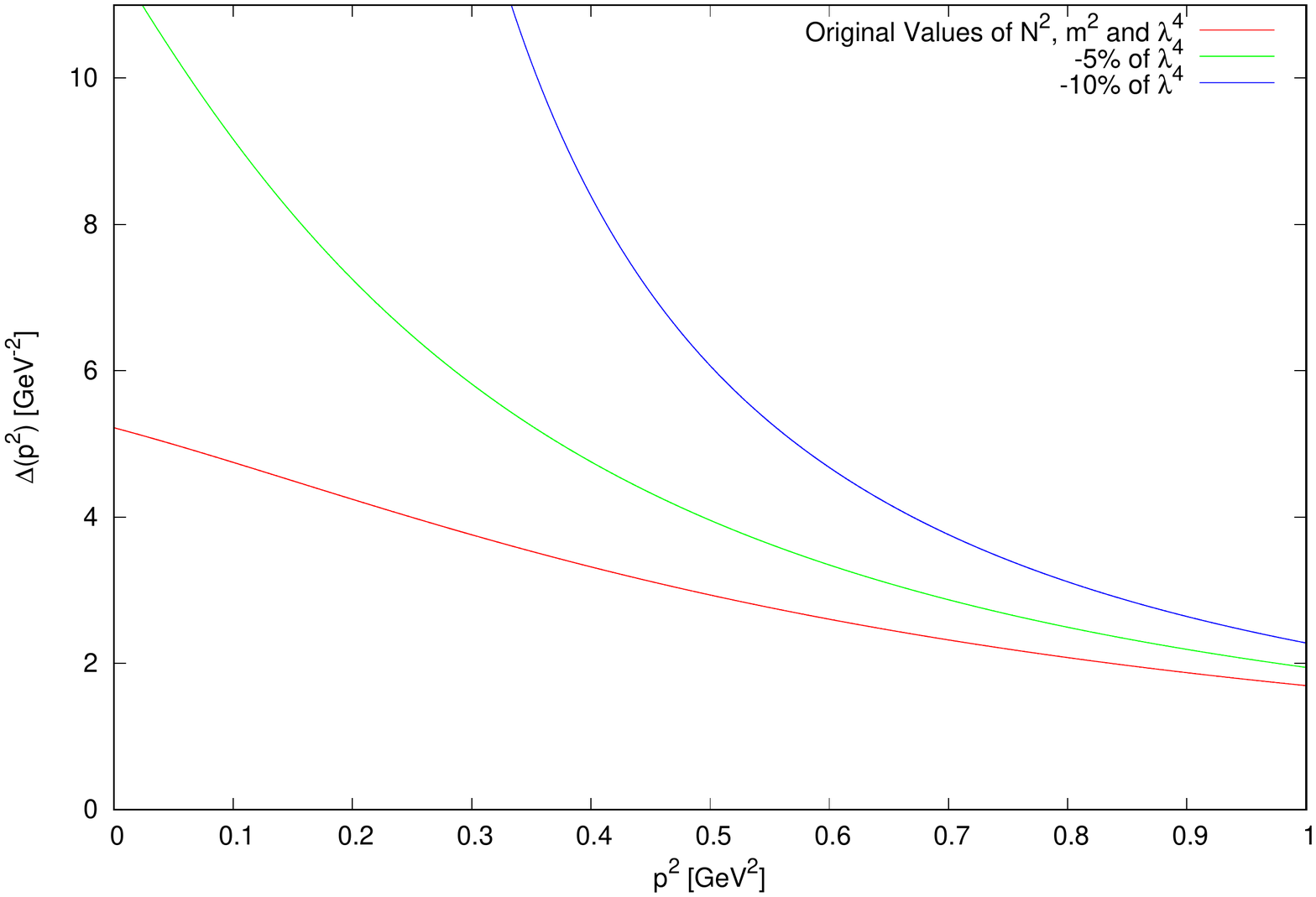}}
\ \subfloat[][]{\includegraphics[width=.35\textwidth]{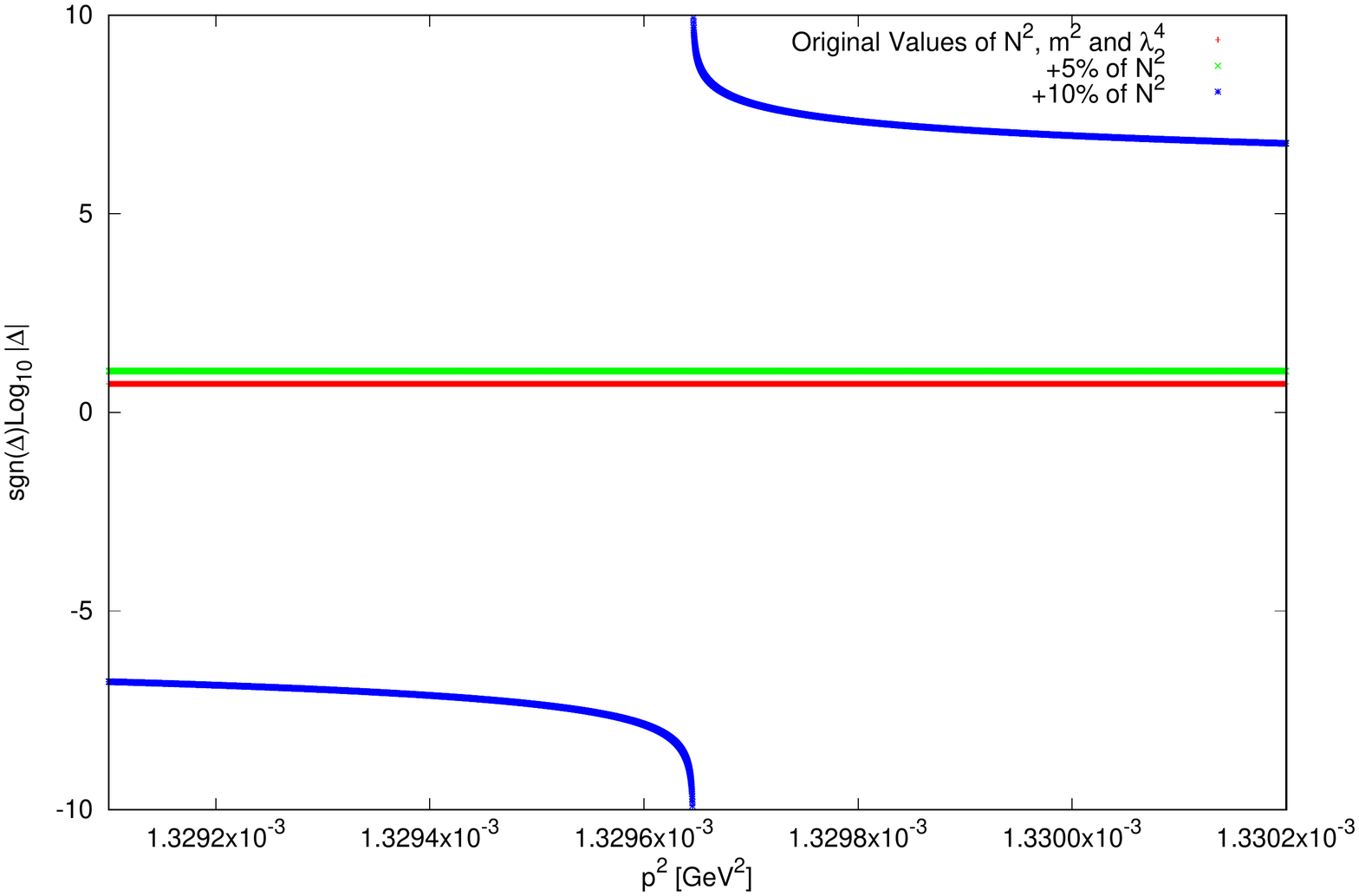}}
\subfloat[][]{\includegraphics[width=.35\textwidth]{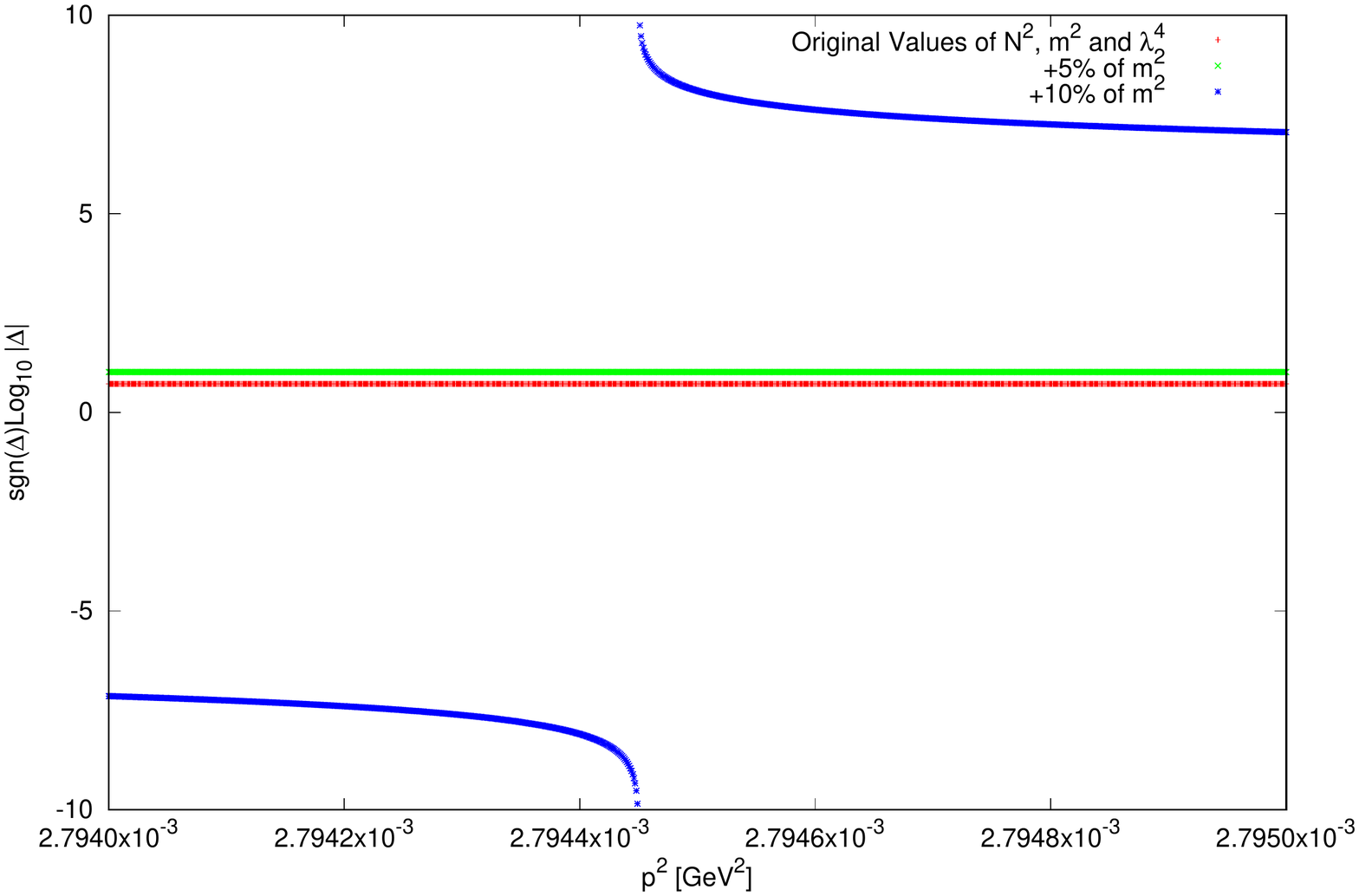}}
\subfloat[][]{\includegraphics[width=.35\textwidth]{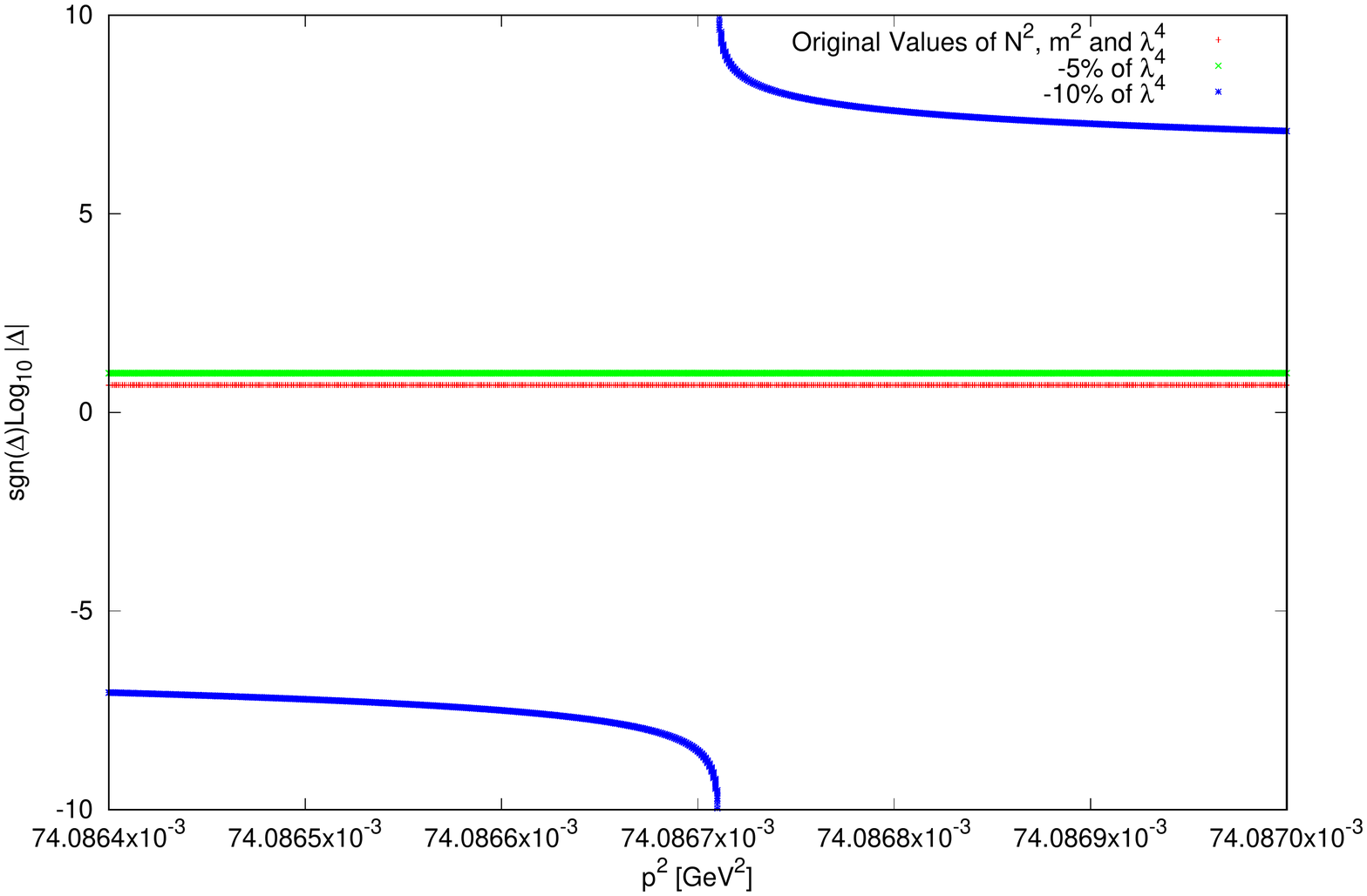}}\caption{Above:
comparison of the curve $\Delta(p^{2})$ in the unit range of square momenta
$p^{2}$ for the original mass lattice values (red line) with respect to (a)
modifying only $N^{2}$ in $+5\%$ (green line) and $+10\%$ (blue line); (b)
modifying only $m^{2}$ in $+5\%$ (green line) and $+10\%$ (blue line); (c)
modifying only $\lambda^{4}$ in $-5\%$ (green line) and $-10\%$ (blue line). Below:
to see more clearly how the modified propagator is out of scale respect to the
lattice values, we plot $\sgn(\Delta)\log_{10}\Delta$ in the square momenta
range close to the real poles which appear (d) in $p^{2}\approx1.3\times
10^{-3}$ $GeV^{2}$ when only $N^{2}$ is modified $+10\%$; (e) in $p^{2}%
\approx2.8\times10^{-3}$ $GeV^{2}$ when only $m^{2}$ is modified $+10\%$; (f) in
$p^{2}\approx7.4\times10^{-2}$ $GeV^{2}$ when only $\lambda^{4}$ is modified
$-10\%$. Observe also how, since the poles are in the real positive axis in
the modified propagators, the function $\Delta(p^{2})$ acquires negatives values.}%
\label{propagator_modif}%
\end{figure}

In this subsection, we consider the non-perturbative gluon EOS taking into
account the appearance of the condensates in the propagator
\cite{vandersickel} which is favored by lattice data \cite{cucchieri1}
(following the same technique to compute the partition function
\eqref{partition_func1}). The RGZ-propagator is \cite{Cucchieri:2011ig}
\begin{equation}
\Delta_{\mu\nu}^{ab}(p)=\delta^{ab}\frac{p^{2}+N^{2}}{p^{4}+p^{2}(N^{2}%
+m^{2})+m^{2}N^{2}+\lambda^{4}}\left(  \delta_{\mu\nu}-\frac{p_{\mu}p_{\nu}%
}{p^{2}}\right)  =\delta^{ab}\Delta(p^{2})\left(  \delta_{\mu\nu}-\frac
{p_{\mu}p_{\nu}}{p^{2}}\right)  \;,\label{propagator_RGZ}%
\end{equation}
$N^{2}$, $m^{2}$ being condensate-related values, and $\lambda$ the Gribov
mass parameter. In this case the \emph{renormalized} RGZ-vacuum energy at
one loop can be written as \cite{polyakov-paper}
\begin{equation}
\mathcal{E}_{v}(T)=\frac{3}{2}(d-1)\sum\limits_{s=-1}^{s=1}\left[
I(T,r_{+}^{2})+I(T,r_{+}^{2})-I(T,N^{2})\right]  -\frac{3}{2}%
I(T,0),\label{vacuum-rgz-renormalized}%
\end{equation}
where the function $I$ is the same as \eqref{I_function_gluon} and $r_{\pm
}^{2}$ are the minus roots of the denominator of the RGZ-propagator, which are
given by
\begin{equation}
r_{\pm}^{2}=\frac{(m^{2}+N^{2})\pm\sqrt{(m^{2}+N^{2})^{2}-4(m^{2}N^{2}%
+\lambda^{4})}}{2}\;.\label{roots_RGZ}%
\end{equation}
where the condensate values $m^{2}$, $N^{2}$ and the Gribov parameter
$\lambda$ were extracted from \cite{Cucchieri:2011ig}:
\begin{eqnarray}\label{latticec}
N^{2} &=& 2.51~\hbox{GeV}^{2}\; , \nonumber\\
m^{2} &=& -1.92~\hbox{GeV}^{2}\; ,   \\
\lambda^{4} &=&5.3~\hbox{GeV}^{4} . \nonumber
\end{eqnarray}
In Figure \ref{pres_energy_temp_RGZ} are plotted the pressure $P$ (red line)
and the energy density (black dots) as functions of temperature $T$. Again a region is
observed where the EOS is not well defined. However, in this
case if we change the fit parameters $\pm10~\%$, then the result is drastically
different. In fact, as we show in Figure \ref{pres_ener_modif_RGZ} (a), for a
modification $+10~\%$ of the RGZ parameters, the critical region reduces
considerably. Also, from Figure \ref{pres_ener_modif_RGZ} (b), we can infer
that the EOS can be defined for $e\geq0$ and for almost all $T$ for a
modification $+10\%$ of the RGZ-parameters.
In order to see how the modified parameters change the behavior of
the propagator respect to the original values, we plot in Figure
\ref{propagator_modif} the function $\Delta(p^{2})$ defined in
\eqref{propagator_RGZ} for different values of the condensates parameters
$N^{2}$, $m^{2}$ and Gribov mass parameter $\lambda$. We can see that the
curves are significantly modified for changes of $5~\%$ and $10~\%$ of these
parameters. Even more, for a change of $10~\%$ (either in the condensates or
Gribov mass parameter) the real poles in the positive axis generate negatives
values of $\Delta(p^{2})$. This means that even if by changing the parameters
it is possible fix the thermodynamic problem, the corresponding propagators
are quite off-scale\footnote{In particular, an intuitive way to realize the lattice propagator is very different from the propagator with real poles
(without thermodynamics pathologies) is due to the difference of the
discriminants, which determines whether the poles are real or complex
conjugated. Such a distance is at least as big as the absolute value of the
discriminant of the lattice propagator itself, which is around $1.6~\hbox{GeV}^{4}$ for the lattice values \eqref{latticec}.} with
respect to the lattice results.

Taking into account that in dimension $d=3$ there exists also the possibility
to refine the GZ approach \cite{GZd3}, which is not the case for $d=2$
\cite{GZd2}, we can perform the same analysis in three dimensions and the
results are qualitatively the same as in dimension $d=4$.

\section{Conclusions and perspectives}
\label{conclusions}

In this paper, the existence and properties of the EOS derived from the
non-perturbative quark and gluon propagators are analyzed at one loop. In
order to reduce as much as possible the number of numerical integrations (and,
correspondingly, the numerical error) the $\zeta$-function regularization
method is used. The present computations are compatible with previous results
found in the reference \cite{quarkNP6}. However, with the $\zeta
$-function regularization method it is possible to disclose very tiny effects
that, otherwise, are difficult to disentangle from the numerical errors.
Indeed, in the case of quarks, the analysis reveals that, depending on the
exact value of the parameters in the fit of the lattice quark propagator, the
pressure as a function of the temperature (at fixed chemical potential) can be a
non-monotonic function of the temperature.

Even if the effect is small, the non-monotonic behavior of the pressure in
this temperature interval implies that the pressure as a function of the energy
density is not single valued (namely, one cannot define an EOS in the usual
sense). By changing the fit parameters of at least $+10~\%$, this feature
almost disappears. The physical explanation (see Appendix \ref{considerations}) of this result is the following: the non-perturbative propagators analyzed
in the present manuscript can have both complex conjugated poles (as suggested
by the lattice data) and real poles. In both cases such propagators violate
positivity (in the former case due to the complex conjugated square masses, in
the latter case due to the fact that there is always a negative residue) and
so they describe confined degrees of freedom. It is true that complex
conjugated poles are more natural to describe confined degrees of freedom (as
first recognized by Gribov and Zwanziger) and are also supported by lattice
data. However, it is interesting to explore what happens when, instead of a
pair of complex conjugated poles, one gets two real poles with opposite
residues. In this case, as we have explicitly shown in this manuscript, there
is no pathological behavior in the thermodynamics (despite the violation of
positivity). The price to pay is, obviously, that it is not easy to interpret
a propagator with two real poles which can be split into two terms (one
with positive and one with negative residue). Such behavior can appear, for
instance, when considering the semi-classical Gribov approach at finite
temperature \cite{nostro7}. Our personal opinion is that complex conjugated
poles are clearly favored and, correspondingly, the pathologies in the
thermodynamical behavior which are observed are related to the fact that some
degrees of freedom (such as light glueballs) - which, unfortunately, are
difficult to describe analytically - are missing at low temperatures.
Nevertheless, we think that it is an interesting observation that one can both
keep alive the violation of positivity and, at the same time, solve the
thermodynamical pathologies mentioned above by using real poles with opposite
residues. For the sake of clarity, in the Appendix \ref{considerations} we
present a concrete example of how the propagator changes when one moves to real poles.

The same analysis has been performed for the non-perturbative gluon and ghost
propagators in the GZ parametrization, and the results are very similar.
Moreover, in this case, the present analysis is able to clearly distinguish
between the scenarios with and without condensates. The reason is that, if one
insists on extracting an EOS in the usual sense, then it is necessary to use the
refined version which includes the condensates (and which is favored by
lattice data). Otherwise, in the case in which the only non-perturbative
parameter is the Gribov mass, not even a $10~\%$ change in its value can
guarantee the existence of an EOS. The reason of this could be traced on the
fact that it is not possible to avoid complex poles in the gluon propagator
changing the Gribov parameter $\lambda^{4}$ for the GZ approach. This is not
the case for the RGZ-approach, where one has more parameters to play with and for which
could be, in principle, obtained real poles (see Appendix \ref{considerations}
for details). However, even in the case of real poles, one of the residues of
the RGZ-propagator has negative sign, violating positivity of the
K{\"a}ll{\'e}n-Lehmann spectrum representation, indicating still some sort of
confinement (see some examples of this in \cite{nostro7,higgs1,higgs2}).

Our interpretation of the present results (both in the quarks case and in the
gluon case) \textit{is not} that it is mandatory to require the fulfillment of
the three requirements mentioned in the previous sections. Rather, it is that
strong quantum effects could be able to violate (the first of) the three
conditions (the \textit{zeroth} requirement of \cite{ruffini} mentioned in the
previous sections). At least in the cases analyzed here, once the condition to
have a well-defined EOS is satisfied, the causality and Le Chatelier
principles are both satisfied. However, the first condition can be violated,
although in small intervals, and this can have a rather big impact on the
physical applications of the non-perturbative equations of state of quarks and
gluons. In particular, the Rhoades-Ruffini bound would not be applicable
anymore and one could have quark stars more massive than expected. Due to the
fact that already a relatively small change in the fit parameters of the
lattice propagators can enforce the consistency conditions of self-gravitating
QCD matter, it is of course of interest to explore the phenomenological
consequences of modifications of such parameters using the mentioned
consistency conditions as guideline.

Still it is possible that the thermodynamics at two-loop computations could have
better behavior for the pressure and entropy as is the case for the massive
Landau-DeWitt action \cite{RSTW-2loop}. However, as for the (R)GZ-approach higher
loops computations are very difficult, it is not easy to see what happens in
the gluonic sector of this model. Nevertheless, we tend to believe this
behavior of the thermodynamical quantities is not just a technical issue but
it is related to the lack of degrees of freedom at low temperatures (confined
phase). After all, we are using propagators which contain a lot of
non-perturbative information (as they come from very precise lattice fits).
It is worth emphasizing that the existence of the EOS is an assumption of
great importance in the astrophysical context. For instance, a great part of the
theory of gravitational hydrostatic equilibrium is based on the assumption
that there is a well-defined functional relation between pressure and energy
density. As there is also evidence supporting the existence of quark stars
(see the review \cite{quarkstar1,quarkstar2}), it is clear that the
present results can be quite relevant in applications.

On the other hand, it is a very concrete possibility that the future lattice
data will confirm the actual values of the fit parameters of the lattice
propagators and from an analytical point of view the thermodynamics quantities have
the same behavior at two or more loops. Indeed, we have also shown that the
propagators with real poles (which are free from thermodynamical pathologies)
are quite off-scale with respect to the lattice data. Thus, it is unlikely
that higher loops effects can take care of such a difference. In such a case,
an EOS (at least in the usual sense) would be unavailable. A non-single-valued
relation between pressure and energy density suggests that there is some
information missing when considering strongly interacting quarks and/or
gluons. In other words, in such situations, some extra physical parameters
able to properly label the equilibrium states are needed (see Appendix
\ref{considerations}). This, of course, is a quite interesting conclusion. We
think that all these issues are worth to be further investigated in the future.

\section{Acknowledgments}

We would like to thanks D.~Dudal, M.~Guimaraes, B.~Mintz, L.~F.~Palhares,
S.~P.~Sorella, M.~Tissier and N.~Wschebor for enlightening discussions and
very useful suggestions. P.~P. thanks the Universidade do Estado do Rio de
Janeiro (UERJ) for the kind hospitality during the development of this work.
This work is supported by Fondecyt Grants 1160137, 1150246 and 1160423. P.~P.
was partially supported from Fondecyt Grant 1140155. The Centro de Estudios
Cient\'{\i}ficos (CECS) is funded by the Chilean Government through the
Centers of Excellence Base Financing Program of Conicyt. A. Z. is also
partially supported by Proyecto Anillo ACT1406.

\appendix

\section{Detailed computation of $I(T,\mu,\alpha^{2})$}

\label{I_computation}

We saw in the paper that to have a manageable expression of $I(T,\mu,\alpha
^{2})$ is very important for the quark and gluonic sector, so we include in
this appendix the detailed computation of this quantity using $\zeta
$-functions in the fermionic and bosonic case.\footnote{Because the computation
for fermions and bosons are very similar, for completeness we keep $\mu$
during the calculation, but at the end only fermions will be considered with
chemical potential.} The definition of the generic $I$ function is
\[
I(T,\mu,\alpha^{2})=\sum\limits_{n=-\infty}^{+\infty}\int\frac{d^{3}p}%
{(2\pi)^{3}}\ln\Lambda^{-2}\left[  p^{2}+(\omega_{n}-i\mu)^{2}+\alpha
^{2}\right]  .
\]
where $\omega_{n}$ are the Matsubara frequencies ($2\pi nT$ in the bosonic
case and $2\pi(n+1)T$ in the fermionic case), $\Lambda$ is a free parameter
which we use to regularize, and $\alpha^{2}$ is a mass parameter which we allow
to acquire a complex value. We can write $I$ as the derivative with respect to some
auxiliary variable $s$ and then taking the limit $s\rightarrow0$:
\begin{equation}
I=\lim_{s\rightarrow0}\frac{\partial}{\partial s}\left(  -T\Lambda^{2s}%
\sum\limits_{n=-\infty}^{+\infty}\int\frac{d^{3}q}{(2\pi)^{3}}\left(
(\omega_{n}-i\mu)^{2}+\alpha^{2}+\vec{q}^{2}\right)  ^{-s}\right)  .
\end{equation}
Defining a new variable $t$ as $|\vec{q}|=t\sqrt{(\omega_{n}-i\mu)^{2}%
+\alpha^{2}}$ and passing to spherical coordinates, we have
\begin{equation}
I=\lim_{s\rightarrow0}\frac{\partial}{\partial s}\left(  -T\Lambda^{2s}%
\frac{4\pi}{(2\pi)^{3}}\sum\limits_{n=-\infty}^{+\infty}\left(  (\omega
_{n}-i\mu)^{2}+\alpha^{2}\right)  ^{\frac{3}{2}-s}\int_{0}^{+\infty}%
dtt^{2}(1+t^{2})^{-s}\right)  .
\end{equation}
Let us focus on the last integral. We can write it as
\[
\int_{0}^{+\infty}dtt^{2}(1+t^{2})^{-s} =\frac{1}{4}\sqrt{\pi}\frac
{\Gamma(s-\frac{3}{2})}{\Gamma(s)},
\]
where we used the $\Gamma$ property $(s-1)\Gamma(s-1)=\Gamma(s)$. Thus, one
gets
\[
I=\lim_{s\rightarrow0}\frac{\partial}{\partial s}\left(  -T\Lambda^{2s}%
\frac{\Gamma(s-\frac{3}{2})}{8\pi^{\frac{3}{2}}\Gamma(s)}\sum
\limits_{n=-\infty}^{+\infty}\left(  (\omega_{n}-i\mu)^{2}+\alpha^{2}\right)
^{\frac{3}{2}-s}\right)  .
\]
Moreover, taking into account the definition of the $\Gamma$ function,
\[
\Gamma(t)=\int_{0}^{+\infty}x^{t-1}e^{-x}dx\ ,
\]
one gets
\[
\Gamma(s-\frac{3}{2})\left(  (\omega_{n}-i\mu)^{2}+\alpha^{2}\right)
^{\frac{3}{2}-s} =\int_{0}^{+\infty}dw w^{s-\frac{5}{2}}e^{-w\left(
(\omega_{n}-i\mu)^{2}+\alpha^{2}\right)  }.
\]
Defining $y=w4\pi^{2}T^{2}$, $v^{2}=\frac{\alpha^{2}}{4\pi^{2}T^{2}}$, and
$c_{\epsilon}=\epsilon/2-\frac{i\mu}{2\pi}$, where $\epsilon=1$ for fermions
and $\epsilon=0$ for bosons, we arrive at
\[
I=\lim_{s\rightarrow0}-\frac{\partial}{\partial s}\left(  \frac{\Lambda
^{2s}T^{4-2s}}{2^{2s}\pi^{2s-\frac{3}{2}}\Gamma(s)}\int_{0}^{+\infty
}dyy^{s-\frac{5}{2}}e^{-yv^{2}}\sum\limits_{n=-\infty}^{+\infty}%
e^{-y(n+c_{\epsilon})^{2}}\right)  .
\]
Let us focus now on the last sum. We can use the Poisson summation formula,
\begin{equation}
\sum\limits_{n=-\infty}^{+\infty}f(x+n)=\sum\limits_{k=-\infty}^{+\infty
}e^{2\pi ikx}\int_{-\infty}^{+\infty}f(x^{\prime-2\pi ikx^{\prime}}dx^{\prime
}.
\end{equation}
In our case $f(x)=e^{-yx^{2}}$ so, completing the square,
\begin{equation}
\int_{-\infty}^{+\infty}e^{-yx^{\prime2}}e^{-2\pi i kx^{\prime}}%
dx^{\prime\frac{-k^{2}\pi^{2}}{y}}\int_{-\infty}^{+\infty}e^{-y(x^{\prime
}+\frac{ik\pi}{y})^{2}}dx^{\prime\frac{-k^{2}\pi^{2}}{y}}\sqrt{\frac{\pi}{y}},
\end{equation}
where in the last equality we used the result $\int_{-\infty}^{+\infty
}e^{-y(x+a)^{2}}dx=\sqrt{\frac{\pi}{y}}$. We arrive at
\begin{align*}
\sum\limits_{n=-\infty}^{+\infty}e^{-y(n+c_{\epsilon})^{2}}  &  =\sqrt
{\frac{\pi}{y}}\left(  1+2\sum\limits_{n=1}^{+\infty}cos(2\pi kc_{\epsilon
})e^{-\frac{k^{2}\pi^{2}}{y}}\right)  .
\end{align*}
Now, we have
\[
I=\lim_{s\rightarrow0}-\frac{\partial}{\partial s}\left(  \frac{\mu
^{2s}T^{4-2s}}{2^{2s}\pi^{2s-\frac{3}{2}}\Gamma(s)}\int_{0}^{+\infty
}dyy^{s-\frac{5}{2}}e^{-yv^{2}}\sqrt{\frac{\pi}{y}}\left(  1+2\sum
\limits_{n=1}^{+\infty}cos(2\pi kc_{\epsilon})e^{-\frac{n^{2}\pi^{2}}{y}%
}\right)  \right)  .
\]
To compute the $n=0$ mode, the integral of the first term is
\begin{align*}
\int_{0}^{+\infty}dyy^{s-\frac{5}{2}}e^{-yv^{2}}\sqrt{\frac{\pi}{y}}
=\frac{\sqrt{\pi}}{(v^{2})^{s-2}}\Gamma(s-2),
\end{align*}
where we used again the definition of the $\Gamma$ function. Now, in order to
compute the $n\neq0$ modes, we can show that
\begin{align*}
\int_{0}^{+\infty}dyy^{s-\frac{5}{2}}\sqrt{\frac{\pi}{y}}2e^{-yv^{2}%
-\frac{n^{2}\pi^{2}}{y}}=2\frac{\sqrt{\pi}}{(v^{2})^{s-2}}\int_{0}^{+\infty
}dzz^{s-3}e^{-z-\frac{n^{2}\pi^{2}v^{2}}{y}}.
\end{align*}
The last integral can be rewritten in terms of the modified Bessel function of the second kind $K_{\nu}$ \cite{bessel_second}:
\[
\int_{0}^{+\infty}dtt^{-\nu-1}e^{-t-\frac{b}{t}}=\frac{2}{b^{\nu/2}}K_{\nu
}(2\sqrt{b}),
\]
so we have
\[
\int_{0}^{+\infty}dyy^{s-\frac{5}{2}}\sqrt{\frac{\pi}{y}}2e^{-yv^{2}%
-\frac{n^{2}\pi^{2}}{y}}=\frac{2^{2}(v^{2})^{2-s}\sqrt{\pi}}{(n^{2}\pi
^{2}v^{2})^{\frac{2-s}{2}}}K_{2-s}(2n\pi\sqrt{v^{2}}).
\]
Defining
\[
I_{n=0}=\lim_{s\rightarrow0}-\frac{\partial}{\partial s}\left(  \frac{\mu
^{2s}T^{4-2s}(v^{2})^{2-s}}{2^{2s}\pi^{2s-2}\Gamma(s)}\Gamma(s-2)\right)  ,
\]
and
\[
I_{n\neq0}=\lim_{s\rightarrow0}-\frac{\partial}{\partial s}\left(  \frac
{\mu^{2s}T^{4-2s}(v^{2})^{\frac{2-s}{2}}}{2^{2s-2}\pi^{s}\Gamma(s)}%
\sum\limits_{n=1}^{+\infty}n^{s-2}K_{2-s}(2n\pi\sqrt{v^{2}})\cos(2\pi
nc_{\epsilon})\right)  ,
\]
we arrive at
\[
I=I_{n=0}+I_{n\neq0}.
\]
Let us compute $I_{n=0}$ first. Using the properties of the $\Gamma$ function
\[
\Gamma(s)=(s-1)\Gamma(s-1)=(s-1)(s-2)\Gamma(s-2),
\]
then
\[
I_{n=0} =\lim_{s\rightarrow0}-\pi^{2}T^{4}v^{4}\frac{\partial}{\partial
s}\left(  \left(  \frac{\Lambda^{-2}T^{2}v^{2}}{2^{-2}\pi^{-2}}\right)
^{-s}\frac{1}{(s-1)(s-2)}\right)  .
\]
Taking into account
\[
\lim_{s\rightarrow0}\frac{\partial}{\partial s}\left(  \frac{\Lambda^{-2}%
T^{2}v^{2}}{2^{-2}\pi^{-2}}\right)  ^{-s} =-\ln\left(  \frac{\alpha^{2}%
}{\Lambda^{2}}\right)  ,
\]
we have
\[
I_{n=0} =\frac{(\alpha^{2})^{2}}{32\pi^{2}}\left(  \ln\left(  \frac{\alpha
^{2}}{\Lambda^{2}}\right)  -\frac{3}{2}\right)  ,
\]
which looks very similar to the standard dimensional regularization procedure
used in quantum field theory at zero temperature \cite{peskin}. Now, in order
to compute $I_{n\neq0}$, we observe that $\Gamma(s)=\frac{1}{s}-\gamma
+\mathcal{O}(s)$ when $s\to0$, which implies $\frac{1}{\Gamma(s)}=s+\gamma
s^{2}+\mathcal{O}(s^{3})$ when $s\to0$. So, in the limit $s\to0$, there only
survives the term which derives from $\frac{1}{\Gamma(s)}$, i.e.,
\[
\label{I_neq_0}I_{n\neq0}=\frac{\alpha^{2}T^{2}}{\pi^{2}}\sum\limits_{n=1}%
^{+\infty}(-1)^{n\epsilon+1}n^{-2}K_{2}(n\frac{\sqrt{\alpha^{2}}}{T}%
)\cosh(n\mu/T).
\]
For the case of fermions,
\begin{equation}
\label{I_function_result}I(T,\mu,\alpha^{2})=\frac{(\alpha^{2})^{2}}{32\pi
^{2}}\left(  \ln\left(  \frac{\alpha^{2}}{\Lambda^{2}}\right)  -\frac{3}%
{2}\right)  + \frac{\alpha^{2}T^{2}}{\pi^{2}}\sum\limits_{n=1}^{+\infty
}(-1)^{n+1}n^{-2}K_{2}(n\frac{\sqrt{\alpha^{2}}}{T})\cosh(n\mu/T),
\end{equation}
while for bosons the result is
\begin{equation}
\label{I_function_gluon_result}I(T,\alpha^{2})=\frac{(\alpha^{2})^{2}}%
{32\pi^{2}}\left(  \ln\left(  \frac{\alpha^{2}}{\Lambda^{2}}\right)  -\frac
{3}{2}\right)  -\frac{\alpha^{2}T^{2}}{\pi^{2}}\sum\limits_{n=1}^{+\infty
}n^{-2}K_{2}(n\frac{\sqrt{\alpha^{2}}}{T}).
\end{equation}

\section{Some considerations on the non-monotonic behavior of $P(T)$}

\label{considerations}

Let us analyze the expression
\[
B=\sum_{i=1}^{4}c_{i}I^{\alpha_{i}},
\]
considered on the quark sector after having determined the cut-off $\Lambda$
requiring $\log Z(0,0)=0$. $B$ is simply the sum of the terms containing the
Bessel functions. For the sake of clarity we will stop at the first one,
$n=1$:
\[
B=T^{2}\frac{18}{\pi^{2}}\left[  {m_{1}}^{2}K_{2}\left(  \frac{\sqrt{{m_{1}%
}^{2}}}{T}\right)  +{m_{4}}^{2}K_{2}\left(  \frac{\sqrt{{m_{4}}^{2}}}%
{T}\right)  +2\Re\left(  {m_{2}}^{2}K_{2}\left(  \frac{\sqrt{{m_{2}}^{2}}}%
{T}\right)  \right)  \right]\; ,
\]
where, eventually, $m_{1}^{2}=0.848,m_{2}^{2}=0.2148+0.0579i,m_{4}^{2}=0.639$.
Ignoring the $T^{2}$ term (it is of no importance to the following) and
developing around $T=0$, $B$ is found to be
\begin{equation*}
B\sim m_{1}^{(3/2)}e^{-m_{1}/T}+m_{4}^{(3/2)}e^{-m_{4}/T}+2\Re\left[
(m_{2}^{2})^{(3/4)}e^{-m_{2}/T}\right]  \ .
\end{equation*}
Thus, with $m_{2}=m_{2R}+i m_{2i}$, one gets
\begin{equation*}
B\sim m_{1}^{(3/2)}e^{-m_{1}/T}+m_{4}^{(3/2)}e^{-m_{4}/T}+2e^{-m_{2R}/T}%
\Re\left[  (m_{2R}+im_{2i})^{(3/2)}e^{-i m_{2i}/T}\right]\; ,
\end{equation*}
and defining $m_{2}=m_{2R}+im_{2i}=\rho_{2}e^{i\phi}$ we end up with
\begin{equation*}
B\sim m_{1}^{(3/2)}e^{-m_{1}/T}+m_{4}^{(3/2)}e^{-m_{4}/T}+2\rho_{2}%
^{(3/2)}e^{-m_{2R}/T}\cos{\left(  \frac{m_{2i}}{T}+\frac{3}{2}\phi\right)  }%
\; .
\end{equation*}
Now, it is clear that, when $T\rightarrow0$, $B$ oscillates due to the presence
of the $\cos$ function, that is to say, because of the complex masses. Thus, if
$m_{2i}$ is zero, $B$ is strictly monotonic. This is exactly what one can get,
for example, changing by $+9~\%$ the parameters. Indeed, in this case, using
$M_{3}=0.214~\hbox{GeV}^{3}$, $m^{2}=0.697~\hbox{GeV}^{2}$, $m_{0}=0.015~\hbox{GeV}$, we get
$\alpha_{1}=0.194,~\alpha_{2}=0.283,~ \alpha_{3}=0.916~\hbox{GeV}^{2}$ see equation
\eqref{poles1}. While, for the gluons, taking $N^{2}=2.74~\hbox{GeV}^{2},
m^{2}=-2.09~\hbox{GeV}^{2}, \lambda^{4}=5.8~\hbox{GeV}^{4}$ in equation \eqref{roots_RGZ},
we get $r_{+}= 0.55,~r_{-}=0.09~GeV^{2}$. Obviously, as the presence of real
poles over complex conjugated poles only depends on the discriminant of
\eqref{roots_RGZ}, one can just vary one of the three parameters keeping fixed
the other two in such a way as to change the sign of the discriminant itself (in
the case of the quark propagator the analysis of the roots is more complicated
as it involves a cubic equation, but, conceptually, a similar scheme can be
applied). However, this is quite beyond the scope of the present work as our
intention was just to emphasize that with real poles (despite the violation of
positivity related to the negative residue) one can solve the above mentioned
pathological behavior of the equation of state $P=P(e)$. The reason is that
the main goal of the present paper is the analysis of the lattice propagators
both of which (quarks and gluons) strongly favor complex conjugated poles.

Is it possible to obtain a strictly monotonic function even in the presence of two
complex conjugate masses? No, in this case it is possible to ensure that $B$
is positive. For instance, reasonable conditions to achieve this are that either $m_{1}<m_{2R}$, or $m_{4}<m_{2R}$. Indeed, in these cases, assuming for
example $m_{4}<m_{2R}$ and $2\left(  \frac{\rho_{2}}{m_{4}}\right)
^{(3/2)}<1$ one gets
\[
B\sim m_{4}^{(3/2)}e^{-m_{4}/T}\left[  1+\left(  \frac{m_{1}}{m_{4}}\right)
^{(3/2)}e^{\frac{-m_{1}+m_{4}}{T}}+2\left(  \frac{\rho_{2}}{m_{4}}\right)
^{(3/2)}e^{\frac{-m_{2R}+m_{4}}{T}}\cos{\left(  \frac{m_{2i}}{T}+\frac{3}%
{2}\phi\right)  }\right]  \geq0\; .
\]
However, there is no possibility to obtain a strictly monotonic function. The
same will be true for the gluonic sector. In this appendix we presented a concrete
example of how the propagator changes when one moves to real poles but, of
course, our main goal is to disclose the thermodynamical pathologies related
to the complex poles supported by the lattice data and the necessity to include in
the thermodynamical description extra degrees of freedom in order to avoid
such pathologies

We conclude that in the presence of complex conjugated poles, the
thermodynamical quantities as pressure and entropy are not well behaved in
some region, as was already pointed out in \cite{benic}. Thus, the above
considerations strongly suggest that some extra physical parameter is
necessary to properly label the equilibrium states. There are many interesting
options such as flavor parameters (in the case of quarks propagator), group
parameters, and so on.

\end{document}